%% file: 0Main.tex
\documentclass[fleqn,usenatbib]{mnras}

\usepackage{newtxtext,newtxmath}
\usepackage[T1]{fontenc}

\DeclareRobustCommand{\VAN}[3]{#2}
\let\VANthebibliography\thebibliography
\def\thebibliography{\DeclareRobustCommand{\VAN}[3]{##3}\VANthebibliography}

\usepackage{graphicx}	% Including figure files
\usepackage{amsmath}	% Advanced maths commands
\usepackage{tabularx}
\usepackage{appendix}
\usepackage{enumitem}
\usepackage[table,x11names]{xcolor}
\usepackage{multirow}
\usepackage{booktabs} 
\newcommand{\comment}[1]{}
\newcommand{\M}{M$_{\odot}$}
\newcommand{\tsfh}{\tau_{\mathrm{SFH}}}
\newcommand{\tstar}{\tau_{*}}
\newcommand{\yocc}{y\mathrm{_{O}^{CC}}}

\newcommand{\yhecc}{y\mathrm{_{He}^{CC}}}
\newcommand{\yncc}{y\mathrm{_{N}^{CC}}}
\newcommand{\yccc}{y\mathrm{_{C}^{CC}}}
\newcommand{\yheagb}{y\mathrm{_{He}^{AGB}}}
\newcommand{\ynagb}{y\mathrm{_{N}^{AGB}}}

\newcommand{\nktext}{NKT13$_{\mathrm{EXT}}$}
\newcommand{\rgal}{$R_{\mathrm{gal}}$}
\newcommand{\zo}{Z_{\mathrm{O}}}
\newcommand{\zosun}{Z_{\mathrm{O}, \odot}}
\newcommand{\zn}{Z_{\mathrm{N}}}
\newcommand{\zc}{Z_{\mathrm{C}}}
\newcommand{\zfe}{Z_{\mathrm{Fe}}}
\newcommand{\dely}{\Delta Y}
\newcommand{\delysun}{\Delta Y_{\odot}}
\newcommand{\yp}{Y_{\mathrm{P}}}
\newcommand{\vice}{{\tt VICE}}
\newcommand{\kms}{{\,{\rm km}\,{\rm s}^{-1}}}
\newcommand{\feh}{[{\rm Fe}/{\rm H}]}

\title[GCE of Helium]{Modeling the Galactic Chemical Evolution of Helium}

\author[M. K. Weller et al.]{
Miqaela K. Weller,$^{1}$\thanks{E-mail: weller.133@osu.edu}
David H. Weinberg,$^{1}$
James W. Johnson$^{1, 2}$
\\
$^{1}$Department of Astronomy and Center for Cosmology and AstroParticle Physics (CCAPP), The Ohio State University, Columbus, OH 43210, USA\\
$^{2}$The Observatories of the Carnegie Institution for Science, 813 Santa Barbara St., Pasadena, CA, 91101
}

\date{Accepted XXX. Received YYY; in original form ZZZ}

\pubyear{2023}

\begin{document}
\label{firstpage}
\pagerange{\pageref{firstpage}--\pageref{lastpage}}
\maketitle

%\begin{abstract}
%We work to understand and constrain the helium yields produced in both massive stars and asymptotic giant branch (AGB) stars. While the AGB yield sets have relatively good agreement with rather short $\sim$ 300 Myr time delays, the massive star yield sets can span up to factors of 2 at all metallicities. With a better understanding of the production of helium, we derive analytical models to describe how helium evolves with time and and compare these to galactic chemical evolution (GCE) models. We find that the analytical and numerical models agree well and the behavior of $\Delta Y$ and its relation with other elements depends on the production mechanisms for those elements. For oxygen, we find a slight increase in $\Delta Y$/$\Delta Z_{\mathrm{O}}$ over time. This is due to the time-delayed AGB contribution of helium whereas oxygen is mainly produced in massive stars and has nearly instantaneous production with very little late time production. Finally, in comparing our results with literature helium observations, we cannot conclusively exclude any of our models with the current uncertainties on the data. Overall, however, it appears that the data prefer lower helium yields, which will likely come from a decrease in the massive star enrichment.
%\end{abstract}

\begin{abstract}
We examine the galactic chemical evolution (GCE) of $^4$He in one-zone and multi-zone models, with particular attention to theoretical predictions of and empirical constraints on IMF-averaged yields.  Published models of massive star winds and core collapse supernovae span a factor of 2 -- 3 in the IMF-averaged $^4$He yield, $\yhecc$.  Published models of intermediate mass, asymptotic giant branch (AGB) stars show better agreement on the IMF-averaged yield, $\yheagb$, and they predict that more than half of this yield comes from stars with $M=4-8 M_\odot$, making AGB $^4$He enrichment rapid compared to Fe enrichment from Type Ia supernovae.  Although our GCE models include many potentially complicating effects, the short enrichment time delay and mild metallicity dependence of the predicted yields makes the results quite simple: across a wide range of metallicity and age, the non-primordial $^4$He mass fraction $\Delta Y = Y-\yp$ is proportional to the abundance of promptly produced $\alpha$-elements such as oxygen, with  $\Delta Y/\zo \approx (\yhecc+\yheagb)/\yocc$.  
Reproducing solar abundances with our fiducial choice of the oxygen yield $\yocc=0.0071$ implies $\yhecc+\yheagb \approx 0.022$, i.e., $0.022M_\odot$ of net $^4$He production per solar mass of star formation.  Our GCE models with this yield normalization are consistent with most available observations, though the implied $\yhecc$ is low compared to most of the published massive star yield models.  More precise measurements of $\Delta Y$ in stars and gas across a wide range of metallicity and [$\alpha$/Fe] ratio could test our models more stringently, either confirming the simple picture suggested by our calculations or revealing surprises in the evolution of the second most abundant element.
\end{abstract}


%%%%%%%%%%%%%%%%%%%%%%%%%%%%%%%%%%%%%%%%%%%%%%%%%%

%%%%%%%%%%%%%%%%% BODY OF PAPER %%%%%%%%%%%%%%%%%%

\input{1Intro}
\input{2TheoreticalYieldCalculations}
\input{3OneZone}
\input{4MultiZone}
\input{5Conclusions}

\section*{Acknowledgements}

We thank Jennifer Johnson, Marc Pinsonneault, Emily Griffith, Fiorenzo Vincenzo, and Chiaki Kobayashi for valuable conversations.  We thank Liam Dubay for sharing his implementation of Gaussian stellar migration in \vice\ ahead of publication.
This work was supported by NSF grants AST-1909841 and AST-2307621 and by STScI award HST-GO-17303.002-A.

%%%%%%%%%%%%%%%%%%%%%%%%%%%%%%%%%%%%%%%%%%%%%%%%%%
\section*{Data Availability}

{\tt VICE} is a publicly available code and the simulations can be
generated using the described parameters. We do provide three files online that can be used to generate the same yield sets as used in this paper. Please see Appendix \ref{appendix: yields} for a more thorough description.



\bibliographystyle{mnras}
\bibliography{0Main}

%%%%%%%%%%%%%%%%%%%%%%%%%%%%%%%%%%%%%%%%%%%%%%%%%%

%%%%%%%%%%%%%%%%% APPENDICES %%%%%%%%%%%%%%%%%%%%%

\begin{appendices}

\input{6Appendix}

\end{appendices}

% Don't change these lines
\bsp	% typesetting comment
\label{lastpage}
\end{document}

%% file: 1Intro.tex
\section{Introduction}

A few minutes after the Big Bang, the lightest elements were formed: hydrogen ($\sim$75\%) and helium ($\sim$25\%), with trace amounts of deuterium, lithium, and beryllium. About 95\% of the helium present in the Universe today came from this period, known as Big Bang Nucleosynthesis (BBN), making it the second most abundant element \citep{Cyburt2016}. The remaining helium was created by stellar nucleosynthesis and distributed to the interstellar medium (ISM) through the winds and core-collapse supernovae (CCSN) of massive stars and the returned envelopes of intermediate mass asymptotic giant branch (AGB) stars. The mass fraction of helium beyond the primordial abundance, $\Delta Y \equiv Y-\yp$, is comparable to the mass fraction $Z$ of all heavier elements combined.  Despite its central role in BBN and stellar nucleosynthesis, the evolution of helium in large galaxies like the Milky Way (MW) has received relatively little attention, in part because $\Delta Y$ is difficult to measure precisely in stars or in gas.  In this paper we review a variety of theoretical studies of helium yields, then incorporate these yields into one-zone and multi-zone models of galactic chemical evolution and compare their predictions to observational constraints.  We focus on $^4$He, reserving study of the rare but important isotope $^3$He for a future paper.

Gas phase measurements of helium require high temperatures like those found in HII regions and planetary nebulae (PNe), and they are subject to a number of systematic uncertainties.  These include weak and blended emission lines in HII regions and spatial stratification of He$^+$ and He$^{++}$ in PNe, making ionization corrections \citep[e.g.,][]{Zanna22} extremely important. Both regions also suffer from density-temperature degeneracies \citep[see, for example,][]{Peimbert2017}, though using the HeI 10830 Å line can reduce the associated uncertainties \citep[see][]{Aver2015}.  Most observational studies of gas-phase helium have focused on low metallicity galaxies with the goal of constraining $\yp$ \citep[e.g.,][]{Izotov2014, Cooke2018, Aver2021, Kurichin2021, Hsyu2020, Matsumoto2022}. This inference typically relies on linearly extrapolating $Y$ vs. O/H to zero metallicity, and we use our models to investigate the validity of this extrapolation.

Because of systematic uncertainties, observational estimates of $\yp$ span a range that is larger than quoted statistical errors, e.g., $\yp = 0.2551 \pm 0.0022$ from \cite{Izotov2014} vs. $\yp=0.243 \pm 0.005$ from \cite{Fernandez2019}.  Here we adopt $\yp= 0.24721 \pm 0.00014$ based on the BBN {\it predicted} value given cosmological parameters constrained by cosmic microwave background (CMB) data (\citealt{Pitrou2021}; see also \citealt{Planck2018}).  This value is in good agreement with the recent observational estimates of \cite{Fernandez2019} and \cite{Aver2021}.  

The initial helium abundance of the Sun can be inferred from evolutionary models that are tightly constrained by the solar luminosity, radius, age, and photospheric abundances, and by helioseismology data \citep[e.g.,][]{Dappen1991, Basu2004}.  We adopt the value $Y_\odot = 0.2703$ from \cite{Asplund2009}, which is higher than the measured {\it photospheric} abundance because of the effects of diffusion and gravitational settling on the latter.  In combination with our adopted $\yp$, this value implies $\delysun = 0.023$.

Stellar spectroscopic measurements of helium require hot stars \citep[e.g.,][]{Nieva12,Elmasl23} and thus probe young stellar populations.  Other stellar helium abundance estimates come from studies of open and globular clusters using a variety of techniques including isochrone fitting (e.g., \citealt{Brogaard2011}), eclipsing binaries (e.g., \citealt{Brogaard2011,Brogaard2012}), spectroscopic measurements in blue horizontal branch stars (e.g., \citealt{Villanova2009,Gratton2013}), and asteroseismic analysis (e.g., \citealt{McKeever19}).  Asteroseismology offers the possibility of stellar helium abundance measurements (e.g., \citealt{Gai2018}) in much larger samples of stars spanning a wide range of chemical composition and age.  One of our goals is to provide theoretical guidance for interpreting such observations.

\citet{Vincenzo2019} carry out detailed investigations of helium chemical evolution in cosmological chemodynamical simulations \citep{Vincenzo2018a,Vincenzo2018b} of three disk galaxies with distinct star formation histories (SFH) and final stellar masses of $1.6-3.3 \times 10^{10} M_\odot$.  They investigate radial gradients and histories of helium abundance in the gas and stellar components, with particular attention to the relation between $Y$ and abundances of other elements including C, N, O, and Fe.  More recently, \citet{Fukushima2024} studied the enrichment of extremely metal-poor galaxies (EMPGs) and the first galaxies using hydrodynamical simulations, exploring He/H-O/H relations as well as Fe/O ratios.  We aim to complement these cosmological simulations with a more detailed comparison of nucleosynthetic yield tables using simpler GCE models that allow tighter control of parameters and more rapid investigation of their impact.  Like \cite{Vincenzo2019} we examine trends of $Y$ with different elements (finding somewhat different results), and we pay particular attention to the roles of metallicity-dependent yields and the time delay of AGB enrichment in governing helium evolution.

We discuss helium yields in Section~\ref{sec: yields}, comparing predictions from multiple studies and detailing the relation between gross and net yields.  In Section~\ref{sec: onezone} we compute helium evolution in one-zone (fully mixed) GCE models with variations in star formation history, star formation efficiency (SFE), and outflow mass loading. We use \vice\ \citep{Johnson2020,Johnson21} for numerical calculations and show that the results can be well but not perfectly approximated by analytic models that treat enrichment as instantaneous and yields as metallicity-independent.  In Section~\ref{sec: multizone} we turn to multi-zone models of the MW disk including stellar radial migration, similar to those of \cite{Johnson21} but with different yield choices and associated outflow prescriptions.  We compare our model results to a variety of observations and suggest future studies that could test our GCE predictions more stringently.  We summarize our findings in Section~\ref{sec: conclusions}. 

%% file: 2TheoreticalYieldCalculations.tex
\section{Theoretical Yield Calculations}
\label{sec: yields}

We define the dimensionless, IMF-averaged helium yields $\yhecc$ and $\yheagb$ to be the net mass of helium produced by massive stars and AGB stars, respectively, per unit mass of star formation. The evolution of helium mass in the ISM then follows  
\begin{equation}
    \label{enrichment}
    \dot{M}_{\mathrm{He}} = \yhecc\dot{M}_{*} + \yheagb\dot{M}_{*} + rY\dot{M_{*}} + \yp\dot{M}_{\mathrm{acc}} - Y\dot{M_{*}} - \eta Y\dot{M_{*}},
\end{equation}
where $\dot{M}_{*}$ is the star formation rate, $\dot{M}_{\mathrm{acc}}$ is the accretion rate, $\eta = \dot{M}_{\mathrm{out}}/\dot{M}_{*}$ is the mass-loading factor of galactic outflows, and $r$ is the recycling parameter denoting the fraction of stellar mass that is returned by the envelopes of stars at their birth abundance. Equation \ref{enrichment} approximates massive star and AGB enrichment as instantaneous, though our numerical methods below account for the time-dependence of AGB enrichment. For a \citet{Kroupa} initial mass fraction (IMF), the recycling factor $r \approx$ 0.4, and we show below that treating recycling as instantaneous is accurate. Equation \ref{enrichment} implicitly assumes that accreted gas has the primordial helium abundance, $\yp$, and ejected gas has the current helium abundance of the ISM, $Y$. There are subtleties to the definition of net yields as discussed below. We refer to helium production by massive stars as $\yhecc$ because of the connection to CCSN, but much of the helium is released in winds before core-collapse. We set the boundary between massive star and AGB enrichment at $m$ = 8 \M. 

\subsection{CCSN and Massive Star Winds}

Helium is, of course, the direct product of hydrogen fusion in stars. However, newly produced helium may be processed into heavier elements before it escapes the star, and it may not be released to the ISM at all if it is part of the collapsing core that forms a neutron star (NS) or black hole (BH) remnant. Massive stars release much of their helium in the form of stellar winds, and they may release more in an eventual supernova explosion, two contributions that are often tabulated separately. Furthermore, much of the helium released is helium that the star was born with and not helium produced during evolution. Because of the large value of $Y$ (versus the much lower mass fractions of heavier elements), particular care is required in distinguishing a star's net helium production from its total helium release. The appropriate definition of net yields depends on how one accounts for the loss terms and recycling in the chemical evolution equations. We choose a definition that is consistent with Equation \ref{enrichment}, and thus also consistent with analytic models like those of \citet[][hereafter WAF]{Weinberg2017} and numerical models like {\tt VICE}. 

Consider a single star of birth mass $m$ that returns mass $m\mathrm{_{He}^{gross}}$ of helium to the ISM through winds and, potentially, a supernova. Of this, a mass $Y(m -m_{\mathrm{rem}})$ is helium the star was born with, where $m_{\mathrm{rem}}$ is the mass of the NS or BH remnant. We define the net helium yield to be
\begin{equation}
    \label{yield}
    m\mathrm{_{He}^{net}} = m\mathrm{_{He}^{gross}} - Y (m - m_{\mathrm{rem}}).
\end{equation}
The star's formation initially removes a mass $Ym$  of helium from the ISM, so if the net yield were zero (i.e., if there was no new helium produced), then the total helium mass lost to the ISM after its evolution would be $Ym - m\mathrm{_{gross}^{He}} = Ym\mathrm{_{rem}}$. 

For computing the IMF-averaged yield (the net yield per solar mass of star formation), we distinguish explosive and wind yields:
\begin{equation}
    \label{frac}
    \yhecc = \frac{\int_{m\mathrm{_{SN}}}^{m\mathrm{_{max}}} [E(m) m\mathrm{_{He, exp}^{gross}} + m\mathrm{_{He, wind}^{gross}} - Y(m - m_{\mathrm{rem}})] \frac{dN}{dm} dm}{\int_{m\mathrm{_{min}}}^{m\mathrm{_{max}}} m \frac{dN}{dm} dm}.
\end{equation}
We take $m_{\mathrm{min}}$ = 0.08 \M\ and $m_{\mathrm{max}}$ = 120 \M\ as the limits of the IMF and $m_{\mathrm{SN}}$ = 8 \M\ as the minimum mass for a supernova progenitor. Following the notation of \citet{Griffith21}, the explodability function $E(m)$ = 1 for progenitors that explode as supernovae and $E(m)$ = 0 for stars that implode to form black holes. Equation \ref{frac} is the equation used by {\tt VICE} for both He and (with the substitution $Y \rightarrow Z_{\mathrm{X}}$) heavier elements. It differs slightly from Equation 3 of \citet{Griffith21}, who erroneously omitted the $m_{\mathrm{rem}}$ term, which is a small correction for elements with mass fraction $Z_{\mathrm{X}}$ $\ll$ 1 but can be substantial for He because $Y \gtrsim$ 0.25. 

\citet[][hereafter LC18]{Chieffi2018} provide files for both gross and net yields, separating wind and explosive contributions, and we have confirmed that these are related by Equation \ref{yield}. \citet[][hereafter NKT13]{NKT13} report total gross yields ejected by supernovae but do not report wind yields, though these can be calculated. We refer the reader to Section 2.1.1 of \citet{Chiaki2020} for a more thorough explanation. The NKT13 yields do not include He synthesized during the supernova explosion itself, but this contribution is small compared to the He synthesized during the progenitor's pre-supernova evolution. NKT13 also do not report yields for progenitors with $m >$ 40 \M, which are presumed to form BHs.

\begin{figure*}
    \centering
    \centerline{\includegraphics[width=0.90\paperwidth]{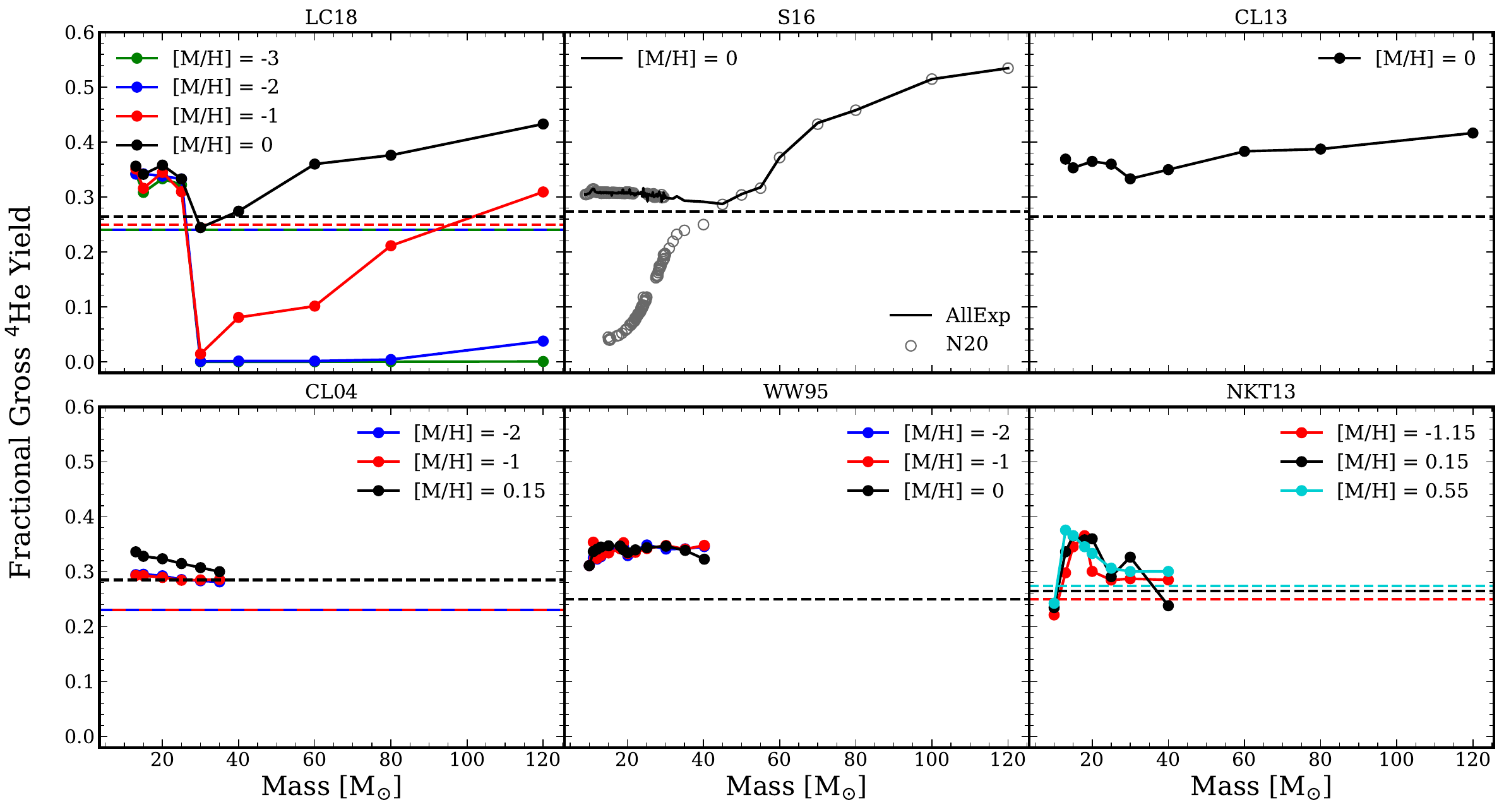}}
    \caption{Fractional gross He yields of massive stars as a function of mass from the stellar evolution and supernova models of (from top left to bottom right) \citet[][LC18]{Chieffi2018}, \citet[][S16]{S16}, \citet[][CL13]{CL13}, \citet[][CL04]{CL04}, \citet[][WW95]{WW95}, \citet[][NKT13]{NKT13}. Most studies report multiple metallicities, and we use the same colors between panels to denote similar metallicities across all of the different studies. Dashed horizontal lines show the birth helium abundance. If the respective color in not shown as a horizontal line, then the birth abundance is the same as for the black dashed case. For WW95, we assumed this value to be 25\%, and for NKT13 we estimate these values based on the other studies. The CL04, WW95, and NKT13 studies do not include winds and do not include progenitors above 40 \M. Black points (AllExp) in the S16 panel show calculations with forced explosions of all progenitors as described by \citet{Griffith21}. Gray open circles (note that this is for [M/H] = 0 as well) show the original S16 results for their N20 central engine, in which many progenitors collapse to black holes. We omit dots for the AllExp models in this panel to aid visibility.}
    \label{fig: ccsngrossbystudy}
\end{figure*}

Figure \ref{fig: ccsngrossbystudy} plots the fractional gross helium yields (i.e., $m\mathrm{_{He}^{gross}}/m$) for a variety of massive star studies. These yields include the sum of both explosions and winds when both are available. At solar metallicity ([M/H] = 0), the LC18 models have gross yields higher than the birth helium abundance at nearly all masses. At lower metallicities, their $m \geq$ 25 \M\ progenitors often have weak or no stellar winds and collapse without explosion, leading to small or zero $m\mathrm{_{He}^{gross}}$. \citet[][hereafter S16]{S16} provide yields at solar metallicity only, but they adopt several different choices of the neutrino-driven central engine, each of which leads to a different ``landscape" of exploding progenitors interspersed with progenitors that collapse to black holes. We adopt the N20 series because it gives a mean CCSN Fe yield in good agreement with empirical estimates \citep{Weinberg2023}. We also show yields from the S16 progenitor models with forced explosions at all masses (AllExp), as reported by \citet{Griffith21}. At $m$ > 40 \M, nearly all of the He is released in winds, so results for N20 and AllExp are indistinguishable even though many of these massive progenitors implode. At lower masses, the imploding progenitors have lower $m\mathrm{_{He}^{gross}}$, though these are still nonzero because of winds. As previously noted, the reported NKT13 yields do not include winds or progenitors with $m >$ 40 \M, but this yield set is commonly used in the literature, so we consider it here. 

For comparison, we also plot in Figure \ref{fig: ccsngrossbystudy} the earlier models from \citet[][hereafter CL04 and CL13, respectively]{CL04, CL13} and \citet{WW95}. CL13 qualitatively agrees with LC18 when they overlap, but the stellar evolution physics in those models has a less sophisticated treatment of stellar parameters such as angular momentum transport and nuclear reaction networks. CL04 and WW95 do not include wind yields and have less complete stellar physics. As a result, we do not consider yields from CL04, CL13, or WW95 in our subsequent calculations.

\begin{figure*}
    \centering
    \centerline{\includegraphics[width=0.90\paperwidth]{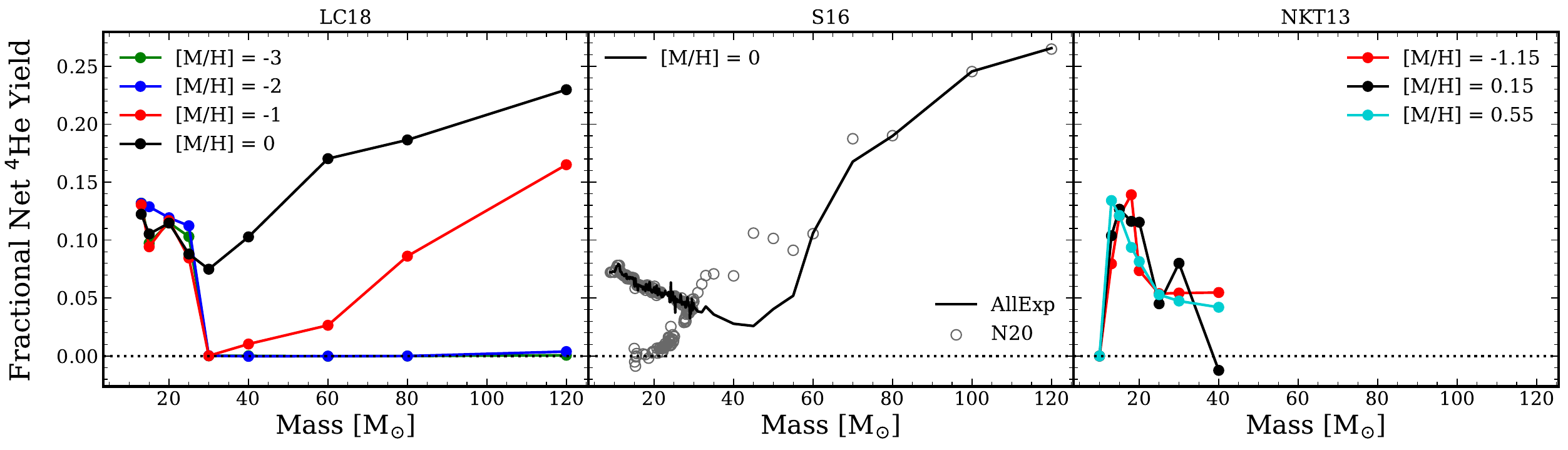}}
    \caption{Similar to Figure \ref{fig: ccsngrossbystudy}, but fractional net massive star yields as a function of mass and metallicity for the four selected stellar evolution studies we will utilize in this paper. The dotted line represents zero helium yield and anything below this line is a net sink of helium, meaning that the star consumed more helium than it produced.}
    \label{fig: ccsnnetbystudy}
\end{figure*}

Figure \ref{fig: ccsnnetbystudy} plots the fractional net yields $m\mathrm{_{He}^{net}}/m$ for the LC18, S16, and NKT13 models, with $m\mathrm{_{He}^{net}}$ defined in Equation \ref{yield}. Note that progenitors with a gross fractional yield lower than $Y$ in Figure \ref{fig: ccsngrossbystudy} may still have a positive net yield because of the $Ym_{\mathrm{rem}}$ term in this definition; the chemical evolution Equation \ref{enrichment} also accounts for the loss of He in these (and other) stars through the $-Y\dot{M_{*}}$ term. The remnant contribution also means that some of the imploding N20 models have higher $m\mathrm{_{He}^{net}}$ than the corresponding AllExp models because the N20 engine has more massive, black hole remnants. For solar metallicity, the LC18, S16/AllExp, and NKT13 models agree at a factor of $\sim$2 level, but not much better. The LC18 and NKT13 models predict weak metallicity dependence of explosive yields, but wind yields in the LC18 models depend strongly on metallicity at high mass due to line driven winds.

For GCE modeling, we care about the IMF-averaged yield, or, the net yield per solar mass of star formation, which is defined by Equation \ref{frac}. For example, to calculate the net yield from LC18, we use Column 8 of Table \ref{tab:lc18} in Appendix \ref{appendix: yields}, which is equivalent to Equation \ref{yield}, then perform the integration (Equation \ref{frac}) where this column reflects the appropriate birth abundance (e.g., $Y$ in Column 2) given [M/H] in Column 1. We compute these for a Kroupa IMF, unless otherwise stated, using {\tt VICE}, though we have also done separate integrations to test the impact of integration schemes given the sparse mass grids available for the models other than S16. We want to integrate from 8 \M\ to 120 \M, but the LC18 and NKT13 grids start at 13 \M\ and 10 \M, respectively. Below these masses, we assume that the fractional net yield $m\mathrm{_{He}^{net}}/m$ is the same as that of the last reported mass in the study. In the S16 models, which start at 9 \M, we find that this procedure gives a reasonably accurate result even if we artificially truncate the table at 13 \M, and it is better than either assuming a constant $m\mathrm{_{He}^{net}}$ or setting the yield to be zero at 8 \M\ and interpolating. On the other hand, because the NKT13 table stops at 40 \M, we have considered both a case in which we set net yields to zero at $m >$ 40 \M\ and an extended model (\nktext) in which we combine the NKT13 yields at $m \leq$ 40 \M\ with the LC18 yields at $m$ > 40 \M. In this high mass regime, stars release helium primarily through winds, which are not reported in the NKT13 calculations, so we consider this extended model to be more realistic.

\begin{figure} 
    \centering
    \includegraphics[width=0.9\columnwidth]{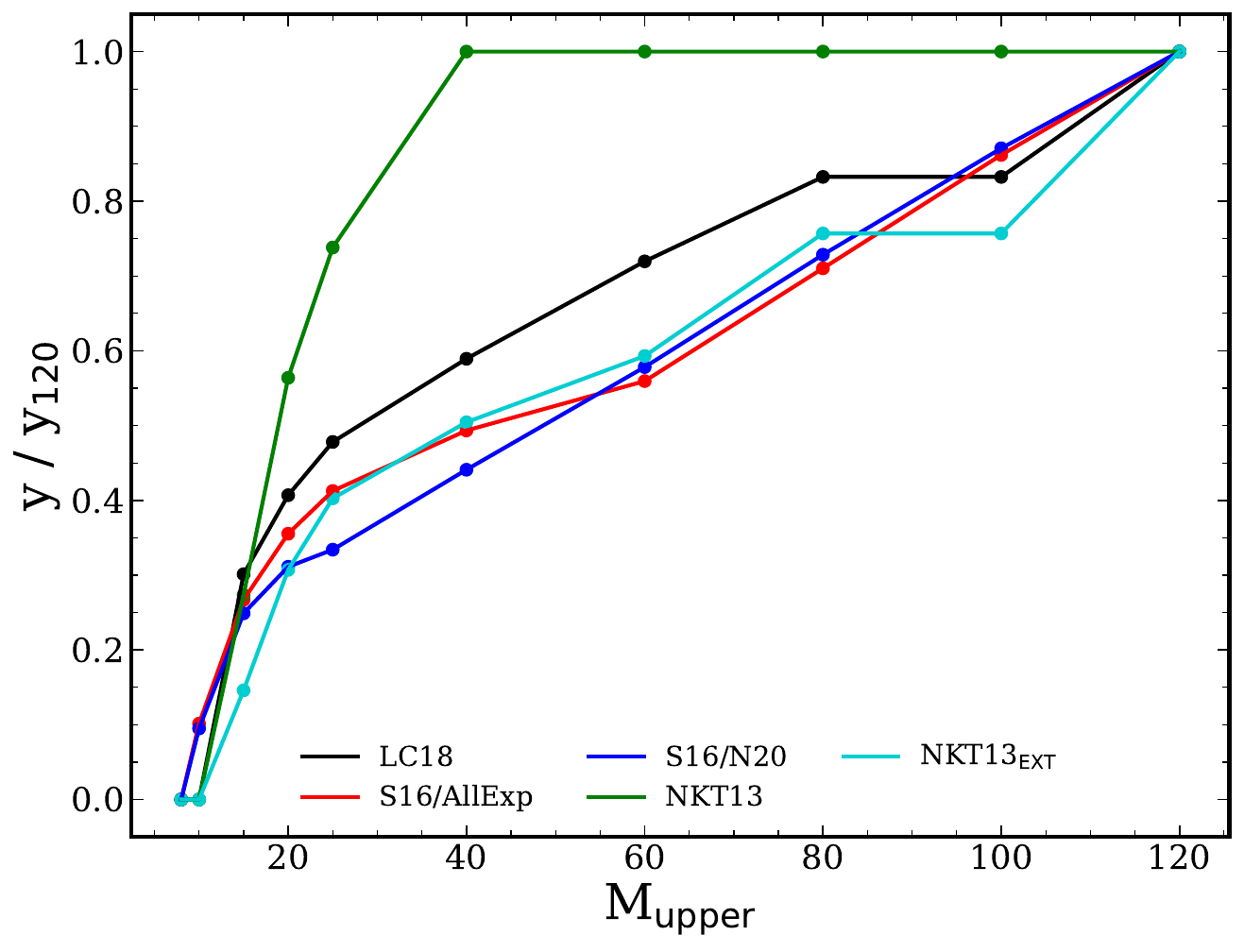}
    \caption{Ratio of the net IMF-averaged yield with different upper mass bounds to the net IMF-averaged yield integrated all the way up to 120 \M\ as a function of mass for solar metallicity. We have included both NKT13 and \nktext, so the solid green line reaches 1.0 at 40 \M. In most cases, about 60\% of all helium from massive stars come from stars with masses less than 60 \M, and there appears to be some agreement between studies. Note that this figure does take into account the initial-final mass relation based on the respective study.}
    \label{fig:fracnetccsn}
\end{figure}

\begin{figure}
    \centering
    \includegraphics[width=0.9\columnwidth]{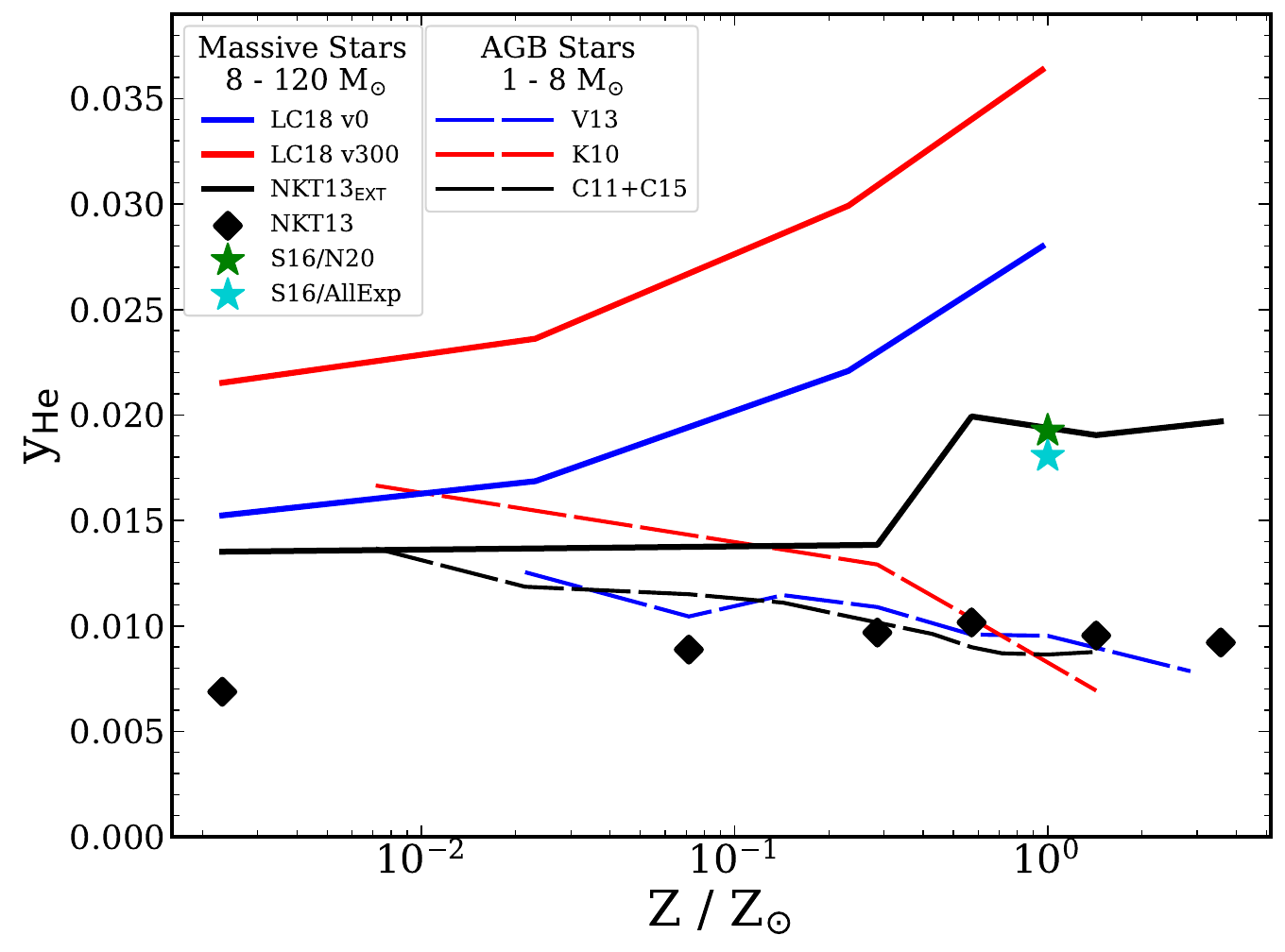}
    \caption{Net IMF-averaged yields for both massive stars (integrated from 8 \M\ to 120 \M) and AGB stars (integrated from 1 \M\ to 8 \M) as a function of metallicity. The S16 study only reports yields at solar metallicity, which is why they only appear as stars here. Despite the S16/AllExp landscape forcing explosions at all progenitor masses, the net yield is slightly lower due to always producing a NS remnant.}
    \label{fig: imf_avg}
\end{figure}

Figure \ref{fig:fracnetccsn} shows how the yield changes if we eliminate contributions from stars with progenitor mass $m > M_{\mathrm{upper}}$ by truncating the integral in the numerator of Equation \ref{frac}. For NKT13, the yield comes entirely from stars with $m \leq$ 40 \M\ by construction, but for the \nktext, LC18, and S16 models, roughly half of the total net yield comes from stars with $m$ = 40 \M\ -- 120 \M. Despite the IMF weighting, the high mass stars are important to the total production of helium. Figure \ref{fig: imf_avg} shows the IMF-averaged net yields as a function of metallicity. Adding these high mass stars increases the IMF-averaged yield of NKT13 by a factor of 1.5 -- 2 at all metallicities (compare black diamonds to the black solid line). Coincidentally but conveniently, the \nktext\ yield agrees perfectly with the S16/N20 yield at solar metallicity. The S16/AllExp yield is slightly (about $\sim$ 5\%) lower, again because of the lower remnant masses and net yield definition. The LC18 yields (blue solid curve) are slightly higher than \nktext\ at low metallicity but nearly 50\% higher at solar metallicity.

In addition to non-rotating progenitors, LC18 compute and report yields for progenitors with surface rotation speeds $v_{\mathrm{rot}} =$ 150 km s$^{-1}$ and $v_{\mathrm{rot}} =$ 300 km s$^{-1}$. Rotationally induced mixing allows more hydrogen to be mixed from the outer regions of the star to regions hot enough for fusion, and it allows more fusion products to be mixed into the envelope where they can be ejected in stellar winds. Both effects increase the helium yield. We show IMF-averaged yields for the $v_{\mathrm{rot}} =$ 300 km s$^{-1}$ models in Figure \ref{fig: imf_avg}, which are a factor of $\sim$1.5 higher than those of the non-rotating models. For our GCE investigations in Sections \ref{sec: onezone} and \ref{sec: multizone}, the \nktext, LC18, and LC18 v300 define a ``low yield," ``medium yield," and ``high yield" scenario for massive star contributions to helium evolution. 

\begin{figure*}
    \centering
    \centerline{\includegraphics[width=0.85\paperwidth]{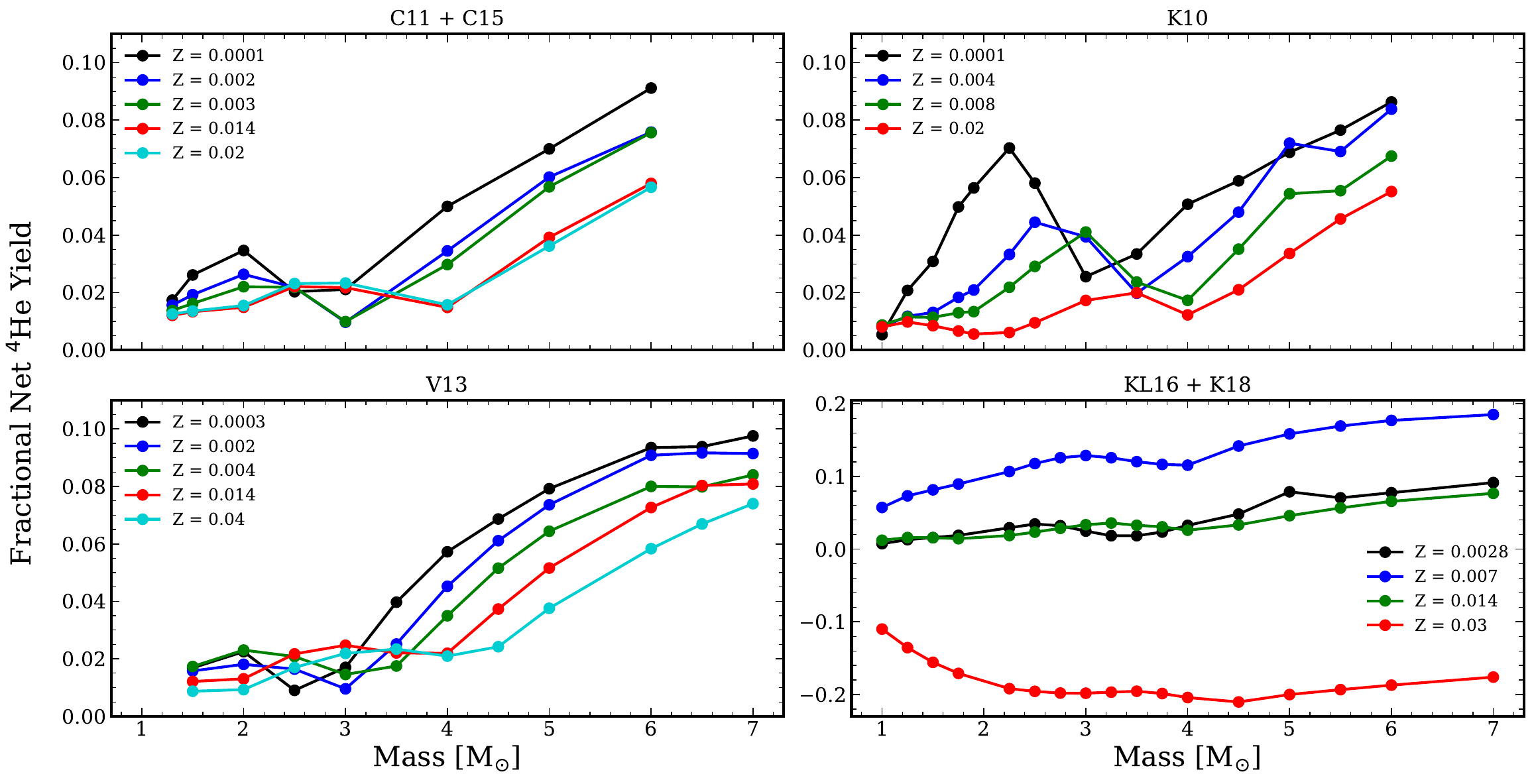}}
    \caption{Fractional net yields for 4 different AGB studies as a function of mass and metallicity. These studies include (from top left to bottom right) \citet{C11, C15}, \citet{K10, V13}, and \citet{KL16, K18} hereafter referred to as the respective panel titles in the figure. While we show KL16 + K18 for comparison, we do not include this study in the rest of the paper, and we note that this study has a different y-axis range than the rest of the panels.}
    \label{fig: agbnetbystudy}
\end{figure*}

\subsection{AGB Stars}

Low mass stars also produce helium during stellar evolution, some of which will enrich the ISM during the AGB phase when the star sheds its outer layer. Birth helium can also be lost to the white dwarf (WD) remnant. Unlike massive star models, there is no separate wind and explosion component, and net yields are typically reported. As such, determining the net helium production in these stars relies less on assumptions than in the previous section, which involved taking into account the type of remnant and respective masses. Equation \ref{yield} still applies, but we need not worry about the right hand side of this equation because studies report the left hand side. We modify Equation \ref{frac} to
\begin{equation}
    \label{frac_agb}
    \yheagb = \frac{\int_{1 {\mathrm{M}_{\odot}}}^{8 {\mathrm{M}_{\odot}}} y_{\mathrm{He}}(m, Z) m \frac{dN}{dm} dm}{\int_{0.08 {\mathrm{M}_{\odot}}}^{120 {\mathrm{M}_{\odot}}} m \frac{dN}{dm} dm}.
\end{equation}
In this equation, $\yheagb(m, Z)$ represents the fractional net yields as a function of mass and metallicity, so we multiply by $m$ to get the yield in solar masses. 

Figure \ref{fig: agbnetbystudy} shows the fractional net helium yields (i.e., $m\mathrm{_{He}^{net}}/m$) as a function of initial mass for different studies and metallicities. \citet[][hereafter C11 + C15]{C11, C15}, \citet[][hereafter K10]{K10}, and \citet[][hereafter V13]{V13} qualitatively agree quite well, but the \citet[][hereafter KL16 + K18]{KL16, K18} studies do not. We note here that the KL16 + K18 yields are run with the {\tt MONASH} stellar evolution code which is known to have problems reproducing empirical trends \citep{Pons2022}, as was also shown for C and N in \citet{Johnson2023}, and the negative yield at high metallicity seems unlikely. As a result, we do not use this study in the rest of this paper. The discrepancies are likely due to a combination of assumptions for convection, mixing, mass loss, opacity, and nuclear reaction rates. All AGB yields have some uncertainty because of these factors. Nonetheless, the remaining three studies are in closer agreement than we found for massive stars. 

\begin{figure}
    \centering
    \includegraphics[width=0.9\columnwidth]{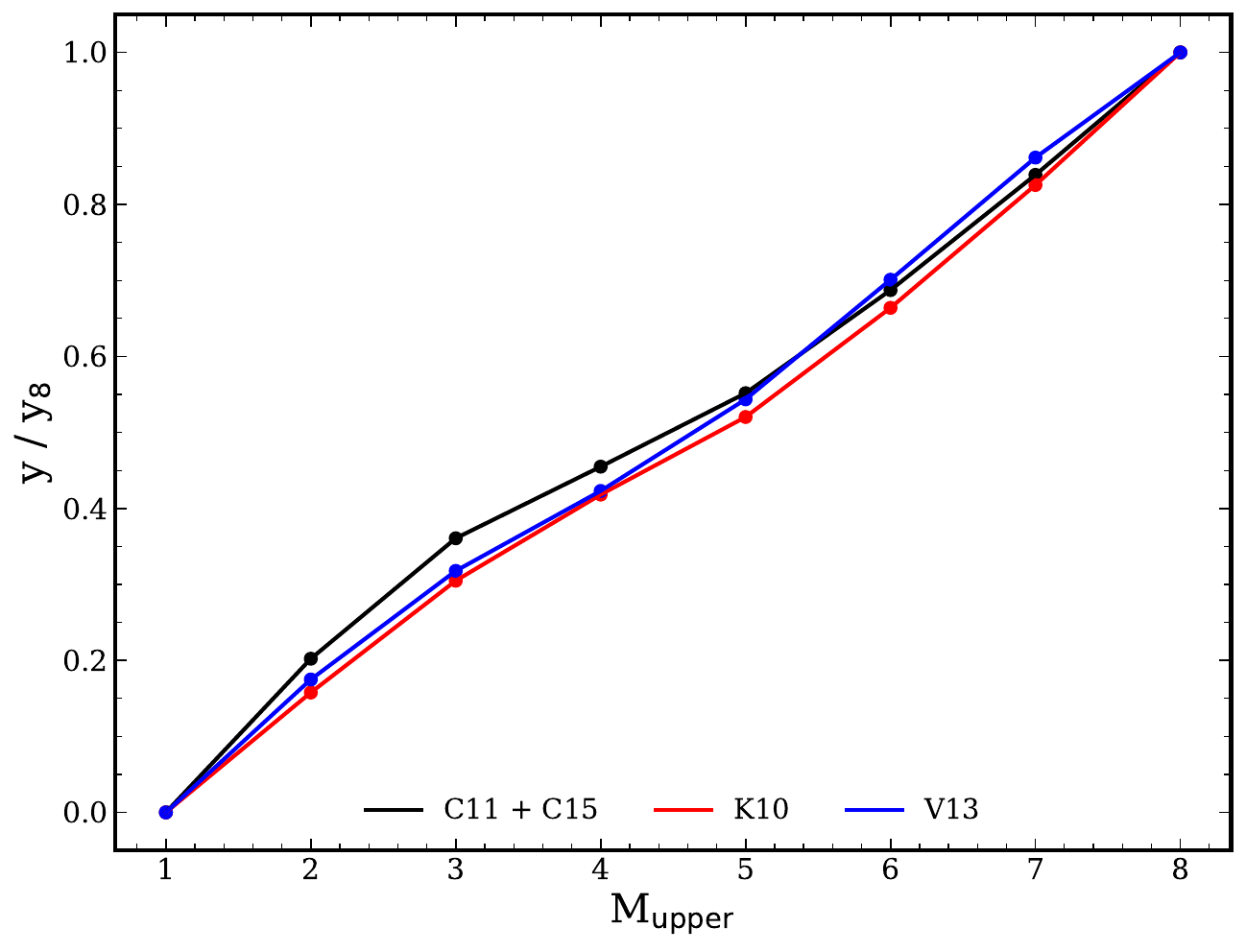}
    \caption{Ratio of the net IMF-averaged yield with different upper mass bounds to the net IMF-averaged yield integrated up to 8 \M. About 60\% of all helium from AGB stars come from stars with masses above 4 \M.}
    \label{fig: agby7}
\end{figure}

While \vice\ computes time-dependent AGB enrichment using mass-dependent lifetimes and yields, our analytic GCE models need the IMF-averaged yield, defined for AGB stars in Equation \ref{frac_agb}. In this low mass regime, we choose to integrate from roughly 1 \M\ to 8 \M. As is reflected in Figure \ref{fig: agbnetbystudy}, the C11 + C15 and K10 tables stop at 6 \M, while the V13 table stops at 7 \M. Here, we use a bi-linear interpolator (in mass and metallicity) for the fractional net yields to extrapolate the yield up to 8 \M. With this in mind, we can make the equivalent of Figure \ref{fig:fracnetccsn} for AGB stars to determine which mass range contributes most to helium production by truncating the integral in the numerator of Equation \ref{frac_agb}. Figure \ref{fig: agby7} plots this cumulative yield fraction, and all studies converge to unity at 8 \M\ due to the bi-linear interpolation scheme despite the tables being cut off at lower masses. We find that, on average, about 60\% of the total helium from AGB stars is produced in stars with $m >$ 4 M$_{\odot}$. The cumulative trend here is nearly linear with mass, a simple result that is consistent among the different studies. Each $\Delta m$ = 1 \M\ contributes about 15\% of the total helium produced from AGB stars. 

\begin{figure*}
    \centering
    \centerline{\includegraphics[width=0.9\paperwidth]{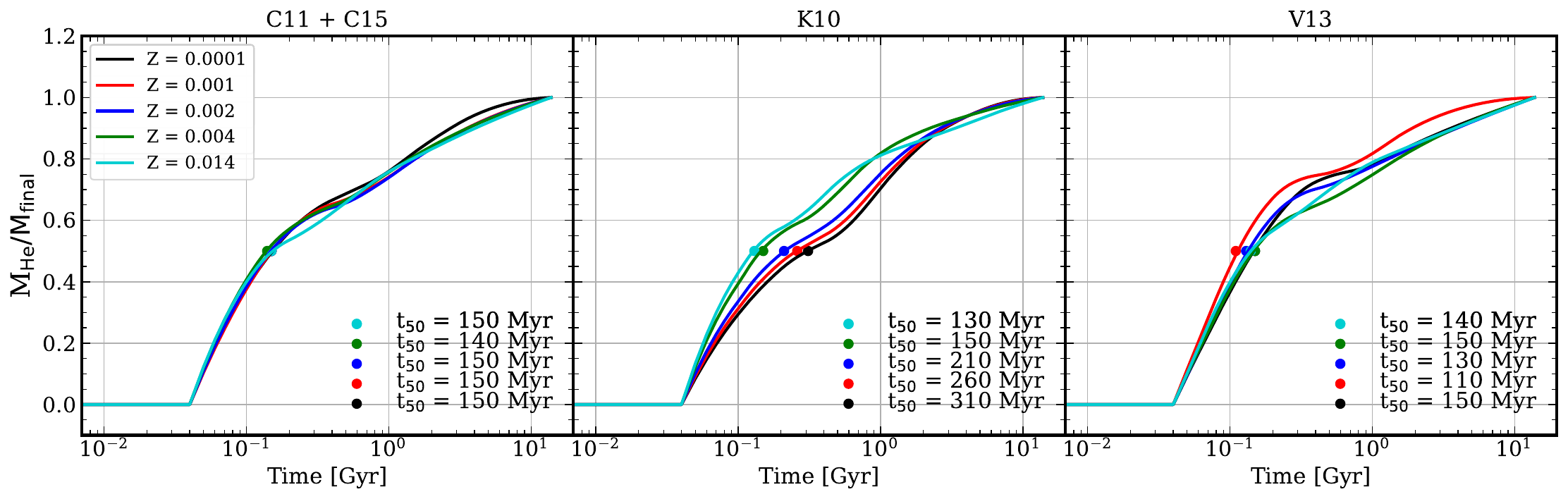}}
    \caption{Delay time distribution of AGB helium enrichment for the three AGB yield studies interpolated to metallicities as labeled. The points represent the time at which half of the total helium is produced, typically about 200 Myr.}
    \label{fig: agbdelay}
\end{figure*}

The He enrichment from AGB stars is delayed relative to massive stars. However, since most of the AGB enrichment comes from stars with $m \geq$ 2 \M, this delay is fairly short. Figure \ref{fig: agbdelay} plots the mass of helium produced by a single stellar population (SSP) as a function of time, normalized by the mass produced down to a 1 \M\  turnoff for different metallicities and AGB studies. We find that about half of the AGB star helium ejection occurs within about 200 Myr. Since the original enrichment equation (Equation \ref{enrichment}) assumes instantaneous production, we should expect deviations from the analytic solutions discussed in the following section. However, instantaneous enrichment is a better approximation for AGB helium than for Fe production by SNIa, which has a characteristic delay time of $\sim$1 Gyr. 

The dashed curves in Figure \ref{fig: imf_avg} show the IMF-averaged AGB yield as a function of metallicity. The C11 + C15 and V13 yields agree well at all metallicities. The K10 yields are similar at solar metallicity and about 20\% higher at low metallicity. All three models predict a helium yield that declines with increasing metallicity, but the effect is relatively mild; for C11 + C15 and V13, the drop between 0.1 $Z_{\odot}$ and $Z_{\odot}$ is about 20\%. This decline likely reflects the decreasing strength of AGB shell flashes, as H-shell burning is cooler in metal-rich stars. However, dredge-up can also play a role due to higher metallicity stars experiencing weaker flashes \citep[e.g.,][]{K02}, thus having more helium end up in the remnant.

\begin{table*}{}
\centering
\caption{Table of IMF-averaged helium yields for both CCSN and AGB contributions at specific interpolated metallicities.}

\label{tab:yield_interp}
\begin{tabularx}{0.57\textwidth}{rrrrrrrr}
\hline
\multicolumn{1}{r|}{\multirow{2}{*}{Studies}} & \multicolumn{7}{c}{$Z/Z_{\odot}$} \\
\multicolumn{1}{r|}{} & 0.01 & 0.1 & 0.3 & 0.7 & 1.0 & 1.5 & 2.0 \\ \hline
\multicolumn{8}{c}{Massive Stars} \\ \hline
LC18 v300 & 0.0222 & 0.0261 & 0.0311 & 0.0349 & 0.0365 & 0.0384 & 0.0397 \\
LC18 v0 & 0.0159 & 0.0186 & 0.0232 & 0.0267 & 0.0282 & 0.0298 & 0.0310 \\
\nktext\ & 0.0136 & 0.0138 & 0.0142 & 0.0198 & 0.0194 & 0.0191 & 0.0193 \\ \hline
\multicolumn{8}{c}{AGB Stars} \\ \hline
V13 & 0.0130 & 0.0109 & 0.0108 & 0.0096 & 0.0095 & 0.0091 & 0.0086 \\
C11 + C15 & 0.0133 & 0.0113 & 0.0102 & 0.0087 & 0.0086 & 0.0088 & 0.0089 \\
K10 & 0.0166 & 0.0154 & 0.0128 & 0.0098 & 0.0086 & 0.0066 & 0.0047 \\ \hline
\end{tabularx}
\end{table*}

\subsection{Summary}

Table \ref{tab:yield_interp} list the IMF-averaged yields of massive stars and AGB stars interpolated by the methods described in this paper to metallicities $Z/Z_{\odot}$ = 0.01, 0.1, 0.3, 0.7, 1.0, 1.5, and 2.0. Yield tables are typically computed at a specified $Z$, which we convert to $Z/Z_{\odot}$ using $Z_{\odot}$ = 0.014. For massive stars, we list yields for LC18 with $v_{\mathrm{rot}} = 0 \kms$ and $v_{\mathrm{rot}}= 300 \kms$ and for \nktext. As shown in Figure \ref{fig: imf_avg}, the three massive star yield sets span nearly a factor of two in $\yhecc$, while the AGB yield sets are in much closer agreement. Total AGB star helium production is generally lower than massive star production, though the AGB and \nktext\ yields are similar at low metallicity. When not otherwise specified, our models in Section \ref{sec: onezone} use the combination of LC18 $v_{\mathrm{rot}}=0$ yields for massive stars and V13 yields for AGB stars. In Section \ref{sec: multizone}, we switch our default massive star yields to \nktext\ because the LC18 yields lead to overproduction of He.

%% file: 3OneZone.tex
\section{One-Zone Galactic Chemical Evolution}
\label{sec: onezone}

We first consider one-zone galactic chemical evolution models (see reviews by \citealt{Tinsley1980} and \citealt{Matteucci21} for discussion), which assume that the star-forming gas reservoir is fully mixed and thus chemically homogeneous at all times. With the further approximations that massive star and AGB enrichment is instantaneous, recycling of birth abundances is instantaneous, and accreted gas has the primordial helium abundance, one arrives at Equation \ref{enrichment} for $\dot{M}_{\mathrm{He}}$, where $M_{\mathrm{He}}$ is the mass of helium in the ISM. The ISM helium mass fraction is $Y$ = $M_{\mathrm{He}}/M_{\mathrm{gas}}$ = $M_{\mathrm{He}}/\left( \tstar \dot{M}_{*} \right)$ where $\tstar \equiv M_{\mathrm{gas}} / \dot{M_{*}}$ is the star formation efficiency (SFE) timescale. For metallicity-independent yields and time-independent values of $\tstar$ and $\eta$, one can use the methods of WAF to find analytical solutions for $Y(t)$ for selected star formation histories. 

We will compare these analytic solutions to numerical solutions, computed with {\tt VICE}, that incorporate time-delayed AGB enrichment, continuous recycling, and metallicity-dependent yields. When scaling metal abundances to solar values, we use the photospheric abundances of \citet{Magg2022} increased by 0.04 dex to account for diffusion (see \citealt{Weinberg2023} for further discussion). As a reference value of the solar birth helium abundance, we adopt $Y_{\odot}$ = 0.2703 from \citet{Asplund2009}, though some studies find higher values of (e.g., $Y_{\odot}$ = 0.278 from \citealt{Aldo2010}). The solar photospheric abundance is significantly lower (about 0.25) today because of diffusion. The change in helium relative to the primordial value is $\dely_{\odot} = Y_{\odot} - \yp = 0.023$ for our adopted values of $Y_{\odot}$ and $\yp$. 

For an exponential star formation history, $\dot{M}_{*} \propto e^{-t/\tsfh}$, the WAF solution for the ISM oxygen mass fraction is
\begin{equation}
    \label{oxy_analytic}
    \zo = Z_{\mathrm{O, eq}} \left(1 - e^{-t/\bar{\tau}}\right)
\end{equation}
with
\begin{equation}
    \label{zoeq}
    Z_{\mathrm{O, eq}} = \frac{\yocc}{1 + \eta - r - \tstar/\tsfh}
\end{equation}
and
\begin{equation}
\label{tbar}
    \bar{\tau} = \frac{\tstar}{1 + \eta - r - \tstar/\tsfh}.
\end{equation}
The abundance approaches an equilibrium value $Z_{\mathrm{O, eq}}$ for $t \gg \bar{\tau}$, a condition that is frequently satisfied for $t$ > a few Gyr. If the initial helium mass fraction is $\yp$ and the accreted gas also has abundance $\yp$, then the corresponding solution for $Y$ is
\begin{equation}
    \label{he_analytic}
    Y = \yp + \dely = \yp + \Delta Y_{\mathrm{eq}} \left( 1 - e^{-t/\bar{\tau}} \right)
\end{equation}
with
\begin{equation}
    \label{yeq}
    \Delta Y_{\mathrm{eq}} = \frac{\yhecc + \yheagb}{1 + \eta - r - \tstar/\tsfh}.
\end{equation}

Conceptually, one can treat the helium synthesized by stars as a separate element that evolves like oxygen and is added to the primordial helium. By approximating the AGB delay time distribution as an exponential, one could adapt the WAF solution for SNIa Fe to model AGB helium enrichment separately from massive star enrichment. We do not do so here because metallicity-dependent yields already affect the accuracy of Equation \ref{he_analytic} more than AGB time delays. Nonetheless, our numerical tests show that Equation \ref{he_analytic} is accurate enough to be conceptually and practically useful.

For numerical solutions, we use {\tt VICE}\footnote{https://vice-astro.readthedocs.io/en/latest/index.html} (Versatile Integrator for Chemical Evolution; \citealt{Johnson2020}; \citealt{Johnson21}). We are building a framework to implement our helium yields and the empirically constrained yields for carbon, nitrogen, magnesium, oxygen, and iron as a module in {\tt VICE}, to update the main branch available on GitHub\footnote{https://github.com/giganano/VICE} that can be installed via {\tt PyPI}\footnote{https://pypi.org/project/vice/}.  This module will be released in the near future, and in the interim we provide code and tables that can be imported to implement the yields adopted in this paper, as described in Appendix~\ref{appendix: yields}.

While the studies summarized in Section \ref{sec: yields} predict yields that are metallicity dependent, it is not clear which metal abundances they should be most sensitive to (e.g., those that dominate the mean molecular weight or those that dominate the opacity). To simplify the interpretation of our models, we choose to tie the yield metallicity dependence to the solar-scaled oxygen abundance, so that assumptions about SNIa enrichment do not affect our results. Specifically, we interpolate our yield tables described above, and {\tt VICE} calculates the yield using a metallicity
\begin{equation}
    Z = 0.014 \left(\zo / Z_{\mathrm{O, \odot}}\right).
    \label{eqn:zoscale}
\end{equation}
Generically, {\tt VICE} calculates this metallicity with all of the solar-scaled elements in the simulation, but we have modified the source code to only consider oxygen, where we take $\zosun = 7.33 \times 10^{-3}$ based on the \citet{Magg2022} photospheric abundance with a +0.04-dex correction for diffusion. For simplicity, we also adopt a metallicity-independent oxygen yield $\yocc$, which we choose to produce $Z_{\mathrm{O, ISM}} \approx \zosun$ at $t$ = 12 Gyr given our other GCE parameters, rather than adopting the yield from the massive star study we are using for helium yields. This approach ensures that model differences in helium evolution are not driven artificially by differences in oxygen evolution combined with metallicity-dependent yields.

Other than yields, the adjustable parameters of our one-zone models are $\tstar$, $\eta$, and $\tsfh$. For molecular gas in the Milky Way and nearby galaxies, $\tstar \approx$ 2 Gyr \citep{Leroy2008, Sun2022}. We adopt this as a reference value and also consider a model with lower star formation efficiency corresponding to $\tstar$ = 6 Gyr. \citet{Weinberg2023} use the observed mean Fe yield of CCSN \citep{Rodriguez2023} to infer $\yocc \approx \zosun \approx 0.0073$. For plausible choices of $\tstar$ and $\tsfh$ tuned to the solar neighborhood, they find that $\eta \approx 0.6$ is required to achieve $Z_{\mathrm{O, eq}} \approx \zosun$ (see Equation \ref{zoeq}). We consider models with $\eta$ = 0.6 and $\eta$ = 0 (no outflows). We adopt $\tsfh = 6$ Gyr for our $\tstar = 2$ Gyr models and a slightly longer $\tsfh = 8$ Gyr for our $\tstar = 6$ Gyr model so that $\bar{\tau}$ does not become overly long. Table \ref{tab:gce_params} lists our three adopted parameter combinations, which we choose to illustrate a range of behaviors, and the resulting evolutionary timescale $\bar{\tau}$ and the adopted oxygen yield $\yocc$. We choose $\yocc$ separately in each model so that $\zo=\zosun$ at $t=12\,{\rm Gyr}$.

\begin{table}{}
\centering
\caption{GCE parameter combinations for $\tstar$, $\tsfh$, $\eta$. We also include the calculated $\yocc$ and $\bar{\tau}$ based on the parameter combinations.}

\label{tab:gce_params}
\begin{tabularx}{0.97\columnwidth}{rrrrrr}
\hline
 & $\tstar$ {[}Gyr{]} & $\tsfh$ {[}Gyr{]} & $\eta$ & $\yocc$ & $\bar{\tau}$ {[}Gyr{]} \\ \hline
Fiducial & 2 & 6 & 0.6 & 0.0064 & 2.31 \\
Zero Outflows & 2 & 6 & 0.0 & 0.0024 & 7.5 \\
Long SFH & 6 & 8 & 0.6 & 0.0056 & 4.44 \\ \hline
\end{tabularx}
\end{table}

\begin{figure}
    \centering
    \centerline{\includegraphics[width=\columnwidth]{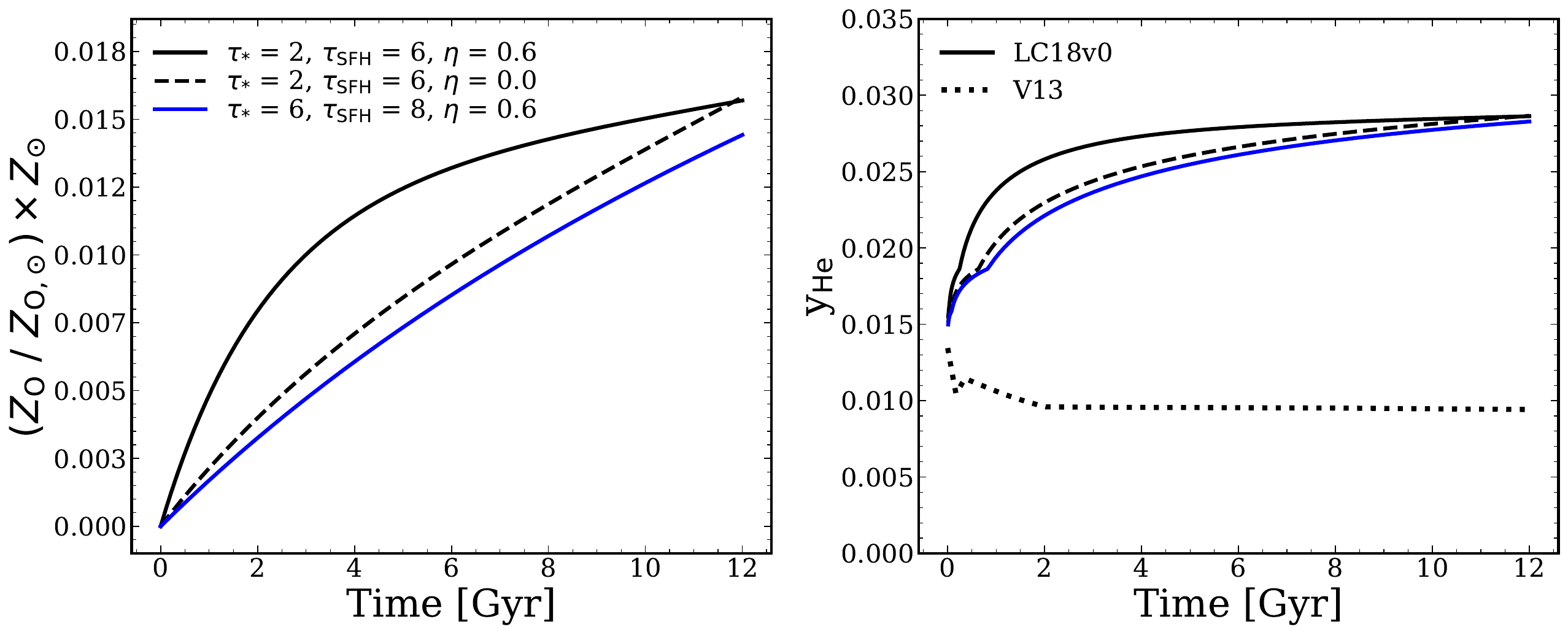}}
    \caption{\textit{Left:} Evolution of the metallicity with an exponentially declining SFH using the metallicity dependent yields of LC18. This is the scaled solar metallicity of oxygen only, as this is what we tie the metallicity dependent helium yields to. \textit{Right:} The IMF-averaged net LC18 and V13 helium yield as a function of time calculated using the metallicities on the left. If an equilibrium abundance is not achieved, then the simulation spends a longer period of time at lower yields.}
    \label{fig: ccsne_metdep_yields}
\end{figure}

The left panel of Figure \ref{fig: ccsne_metdep_yields} shows the evolution of the oxygen abundance in our three one-zone models. The ($\tstar$, $\eta$) = (2, 0.6) model is approaching equilibrium at t > 8 Gyr, but the other two models show a steady rise in $\zo(t)$. The models presented in WAF typically show a clearer equilibrium behavior because WAF used higher $\yocc$ and correspondingly higher $\eta$, leading to shorter evolutionary timescales $\bar{\tau}$. The right panel of Figure \ref{fig: ccsne_metdep_yields} shows the corresponding helium yields from the LC18 $v_{\mathrm{rot}}$ = 0 massive star models with the V13 AGB models, computed at the metallicity implied by $\zo(t)$. 

\begin{figure*}
    \centering
    \centerline{\includegraphics[width=0.80\paperwidth]{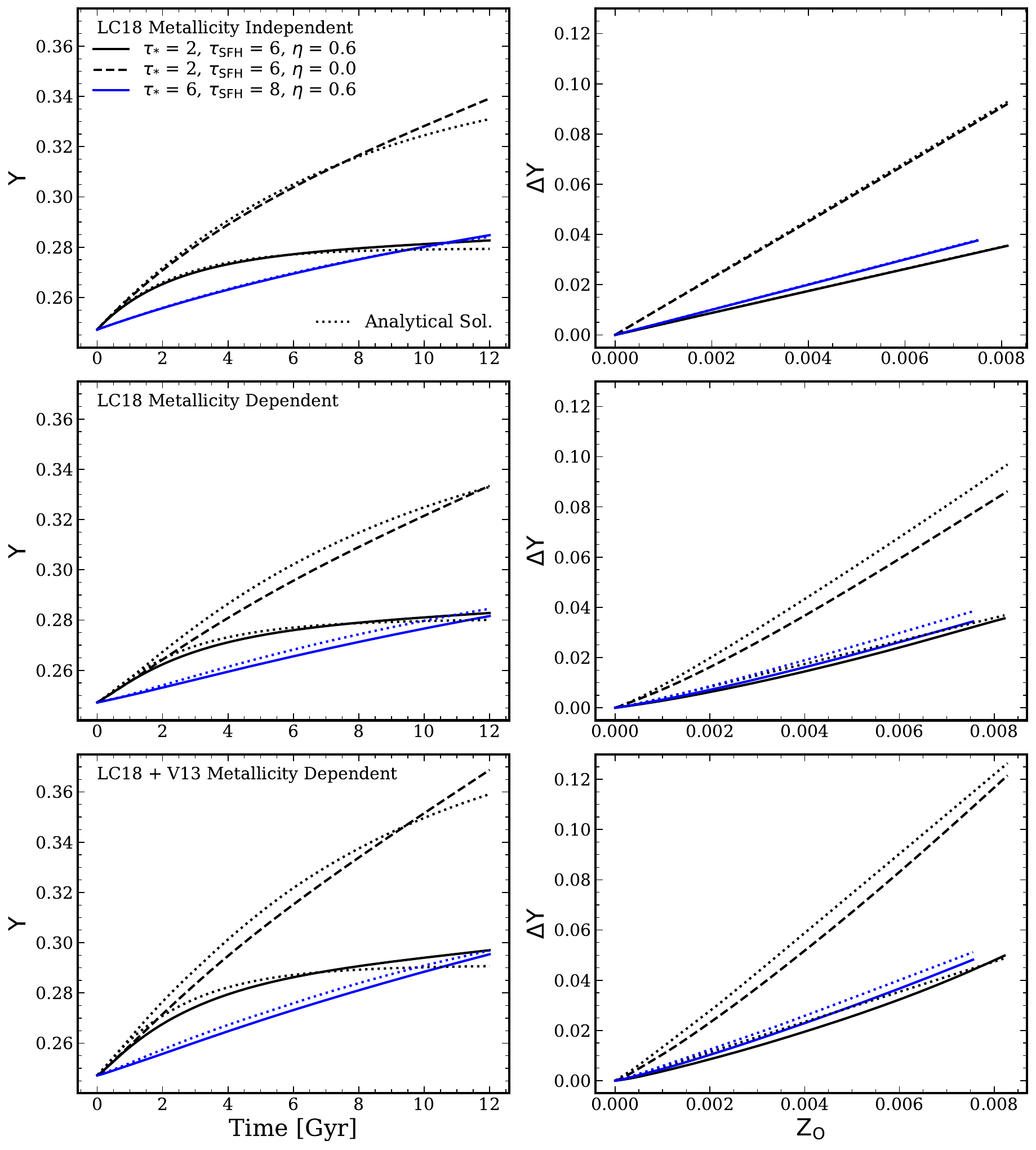}}
    \caption{Evolution of helium in one-zone GCE models. Left panels show $Y$ vs. $t$ and right panels show $\dely$ vs. $\zo$. Solid black, dashed black, and solid blue curves show parameter combinations ($\tstar$, $\tsfh$, $\eta$) = (2, 6, 0.6), (2, 6, 0.0), and (6, 8, 0.6), respectively. \textit{Top:} Models with the metallicity-independent, massive star yields only. \textit{Middle:} Models with the metallicity-dependent LC18 yields but no AGB yields. \textit{Bottom:} Models with metallicity-dependent LC18 massive star yields and V13 AGB yields. In every panel, dotted curves show the analytic model of Equations \ref{oxy_analytic}--\ref{yeq}, with $\Delta Y_{\mathrm{eq}}$ (Equation \ref{yeq}) computed using the value of $\yhecc$ and $\yheagb$ at the metallicity corresponding to $\zo(t)$. The analytic solutions remain accurate even in models that incorporate time-delayed AGB enrichment, metallicity-dependent yields, and continuous recycling. The value of $\yocc$ is adjusted in each model to produce $\zo \approx \zosun$ at $t$ = 12 Gyr (Table \ref{tab:gce_params}).}
    \label{fig: he_onezone}
\end{figure*}

\begin{figure*}
    \centering
    \centerline{\includegraphics[width=0.7\paperwidth]{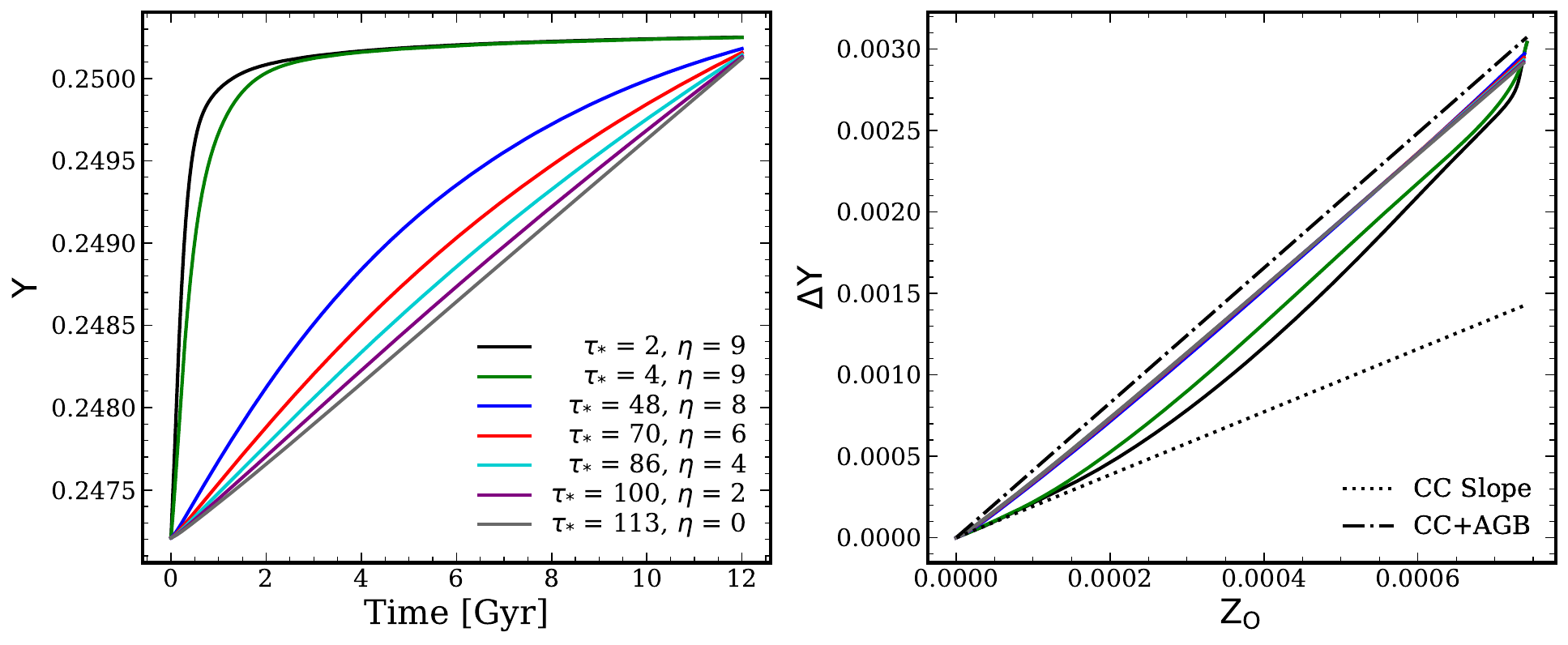}}
    \caption{Similar to Figure \ref{fig: he_onezone} (bottom), but with model parameters chosen to produce $\zo = 0.1 Z_{\odot}$ at $t$ = 12 Gyr. All models have a constant SFR ($\tsfh = \infty$). In the right panel, dotted and dot-dashed lines show the slopes $\dely / \zo\ = \yhecc / \yocc$ and $(\yhecc + \yheagb) / \yocc$, respectively.}
    \label{fig: he_lowmet}
\end{figure*}

Figure \ref{fig: he_onezone} presents the helium evolution of our one-zone models. In the top row, we consider only massive star yields, and we suppress the metallicity dependence of the yield, adopting the $Z = Z_{\odot}$ value of the $\yhecc$ = 0.028 at all metallicities. In this case, the {\tt VICE} models satisfy all the conditions of the analytic solutions except for instantaneous recycling of stellar envelopes. Not surprisingly, the analytic solutions (dotted curves) lie very close to the numerical results. The {\tt VICE} helium abundances are slightly higher at late times because continuous recycling provides a source of helium from the birth abundances of long-lived stars. 

From Equations \ref{oxy_analytic} and \ref{he_analytic}, one can see that the analytic solutions predict
\begin{equation}
    \label{slope}
    \frac{\dely}{\zo} \approx \frac{y_{\mathrm{He}}}{\yocc}.
\end{equation}
For the metallicity-independent yield models, this approximation is near-perfect, as shown in the top right panel of Figure \ref{fig: he_onezone}. The different slopes for the $\dely - \zo$ relations reflect the different $\yocc$ in the three models.

The middle row of Figure \ref{fig: he_onezone} shows results for the metallicity-dependent LC18 yields. For the analytic solution, we retain Equation \ref{he_analytic} and compute $\Delta Y_{\mathrm{eq}}$ at each time using the yield for the current metallicity $\zo(t)$ as shown in Figures~\ref{fig: imf_avg} and~\ref{fig: ccsne_metdep_yields}. The analytic solutions now overpredict the numerical results by a modest factor because the helium yield increases with metallicity, and the ISM includes gas enriched by lower metallicity stars with lower helium yields. The predicted $\dely - \zo$ tracks are now curved because of the yield metallicity dependence. However, the differences between the upper and middle panels are small compared to the impact of the GCE parameters and oxygen yield.

The bottom panels of Figure \ref{fig: he_onezone} present the main result of this Section, the helium evolution of models including metallicity-dependent massive star enrichment and metallicity and time-dependent AGB enrichment, using the LC18 $v_{\mathrm{rot}}$ = 0 and V13 yields. The shapes of $Y(t)$ and $\dely$ vs. $\zo$ are similar to the previous cases, but the helium abundance is higher because of the addition of AGB yields. For the analytic solution, we now use the sum of massive star and AGB yields in Equation \ref{yeq}, again evaluating them at the metallicity $\zo(t)$. Although the analytic solutions treat AGB enrichment as instantaneous, they reproduce the numerical results more accurately than they did in the middle row because the sum of $\yhecc$ and $\yheagb$ has weaker metallicity dependence than $\yhecc$ alone.

The accuracy of the analytic model implies that Equation \ref{slope} is a useful guide for inferring the total (massive star plus AGB) helium yield near solar metallicity. Scaling to our fiducial values of $\delysun$ and $\zosun$ and the empirical oxygen yield advocated by \citet{Weinberg2023} gives
\begin{equation}
    \label{he_yield_scaled}
    y\mathrm{_{He}} = 0.022 \left(\frac{\delysun}{0.023}\right) \left(\frac{\zosun}{0.0073}\right)^{-1} \left(\frac{\yocc}{0.0071}\right).
\end{equation}
\citet{Weinberg2023} estimate the uncertainty in the ratio $\yocc / \zosun$ at $\pm$ 0.1-dex, with the largest contribution to the uncertainty coming from the [O/Fe] ratio that corresponds to pure CCSN enrichment. Adopting the \citet{Aldo2010} value of $Y_{\odot}$ = 0.278 would increase $\delysun$ by 35\%, to 0.031. The agreement among the V13, C11 + C15, and K10 AGB yields suggests that the theoretical prediction $\yheagb \approx$ 0.009 at $Z \approx Z_{\odot}$ is robust.

For the fiducial scaling of Equation \ref{he_yield_scaled}, the implied $\yhecc$ = 0.013, which is below the S16 and \nktext\ predictions and well below the LC18 predictions. Increasing $\yocc$ by 0.1 dex raises the implied $\yhecc$ to 0.019, in good agreement with S16 and \nktext. Additionally, increasing $\delysun$ by 35\% raises the implied $\yhecc$ to 0.028, in good agreement with LC18 $v_{\mathrm{rot}}$ = 0. Unfortunately, the uncertainties are large enough to prevent a strong test of massive star yields, but they could be reduced in future studies, and the high $\yhecc$ = 0.037 predicted by the rapidly rotating LC18 models at solar metallicity appears unlikely.

Studies of the primordial helium abundance usually target low metallicity dwarf galaxies (e.g., \citealt{Aver2021}), so the $\dely - \zo$ relation in the regime $\zo \sim 0.1 \zosun$ is of particular importance. While the models illustrated in Figure \ref{fig: he_onezone} traverse this low metallicity regime at early times, dwarf galaxies have low metallicities at the present day, which is usually attributed to a combination of high outflow rates (large $\eta$) and low star formation efficiencies (large $\tstar$) \citep[see, e.g.,][]{Lanfranchi2006, Kirby11, Chisholm2017, McQuinn2019, Sandford22, Johnson23b}. To explore this regime, therefore, we construct one-zone models with combinations of $\eta$ and $\tstar$ that yield $\zo = 0.1 \zosun$ at $t$ = 12 Gyr. Because the gas-rich galaxies observed for primordial He have young stellar populations, we adopt the $\tsfh \rightarrow \infty$ limit that corresponds to constant SFR, implying an evolutionary timescale $\bar\tau = \tstar / (1 + \eta - r)$ (Equations \ref{oxy_analytic} -- \ref{he_analytic}). We choose two high-efficiency models with $\tstar = 2$ and 4 Gyr, both of which require $\eta \approx 9$ to yield $\zo \sim Z_{\mathrm{O, eq}} \sim 0.1 \zosun$ given our adopted $\yocc$. We also examine models with lower $\eta$ = 8, 6, 4, 2, 0, choosing much longer $\tstar$ values so that the models remain below $Z_{\mathrm{O, eq}}$ at $t$ = 12 Gyr. We evolve these models with {\tt VICE} using continuous recycling and the metallicity-dependent helium yields from LC18 (for $v_{\mathrm{rot}} = 0$) and V13.

Figure \ref{fig: he_lowmet} shows $Y(t)$ and  $\dely - \zo$ for these models. The two short-$\tstar$ models rise rapidly to the equilibrium helium abundance implied by Equation \ref{yeq}; because of the large $\eta$, the evolutionary timescale $\bar\tau$ is only $\approx$ 0.2 Gyr and 0.4 Gyr for these models. However, the $Y(t)$ curves for the low-efficiency models are much shallower, approaching a linear rise for long $\tstar$. For these models, $\dely$ vs. $\zo$ is almost perfectly described by a linear trend with a slope $(\yheagb + \yhecc) / \yocc$, computed from the yields at the current metallicity $\zo(t)$. For the high-efficiency models, the trend at the lowest metallicity follows the shallower, CCSN-only slope $\yhecc / \yocc$ before steepening as AGB enrichment becomes significant. All models should exhibit this change of slope, but because the early metallicity evolution follows $\zo = \yocc\ t / \tstar$, the low-efficiency models are still at extremely low $\zo$ when AGB enrichment turns on. The high-efficiency models also show a small uptick in $\dely$ near the final values of $\zo$. This uptick arises because low mass AGB stars continue to add helium after $\zo$ saturates at equilibrium. While these subtle changes of slope mark interesting physical transitions, our results suggest that a linear $\dely - \zo$ relation should be accurate for modeling helium abundances of low metallicity galaxies because the predicted time delays of enrichment are short even when considering AGB contributions, and the predicted metallicity dependence of $\yheagb + \yhecc$ is weak over the range $0 < Z < 0.1 Z_{\odot}$.

%% file: 4MultiZone.tex
\section{Multi-zone Galactic Chemical Evolution}
\label{sec: multizone}

We now turn to multi-zone models to predict the evolution of helium throughout the MW disk. We again use {\tt VICE} for these calculations, closely following the approach of \citet{Johnson21} but with some significant parameter changes. {\tt VICE} tracks the evolution of gas-phase abundances in a series of disk annuli, each of which can have a different star-formation history and outflow mass-loading $\eta$. The SFE timescale $\tstar$ is determined locally from the gas surface density based on an emperical star-formation law. Stars formed at a given radius and time inherit the corresponding gas phase abundances, and CCSN enrichment is contributed immediately to the same annulus. However, {\tt VICE} incorporates radial migration of stars \citep{Binney2000, Schonrich2009}, and a stellar population's delayed enrichment from SNIa, AGB, and continuous recycling are distributed over time following that population's radial path. \citet{Johnson21} show that this model can reproduce many observed features of stellar abundance and age distributions as a function of radius (\rgal) and midplane distance ($\lvert z\rvert$), though it does not produce an [$\alpha$/Fe] bimodality as sharp as that observed in the MW disk.

All of the helium yield tables discussed in Section \ref{sec: yields} will eventually be incorporated into the public version of {\tt VICE}\footnote{https://github.com/giganano/VICE}. In some cases, this will require corrections to properly account for definitions of net yields. We adopt the same, empirically motivated, ``inside-out" star-formation history and star-formation law as \citet{Johnson21}. To reproduce the observed Galactic [O/H] gradient, \citet{Johnson21} adopt
\begin{equation}
    \label{eta}
    \eta (R_{\mathrm{gal}}) = \frac{y_{\alpha}^{\mathrm{CC}}}{Z_{\alpha, \odot}} 10^{(0.08 \mathrm{kpc}^{-1})(R_{\mathrm{gal}} - 4 \mathrm{kpc}) - 0.3} + r - 1.
\end{equation}
For a constant SFR, this formula leads to an equilibrium abundance [O/H] that is +0.3 at \rgal\ = 4 kpc with a gradient of -0.08 dex kpc$^{-1}$, as shown by the $\tsfh \rightarrow \infty$ limit of Equation \ref{zoeq}. For the oxygen yield, \citet{Johnson21} adopt $\yocc / \zosun = 0.015 / 0.0057 \approx 2.6$, for which Equation \ref{eta} implies $\eta \approx$ 1.9 at the solar radius \rgal\ = 8 kpc.

As discussed in Section \ref{sec: onezone} (cf. Equation \ref{he_yield_scaled}), for the yield $\yocc \approx \zosun$ advocated by \citet{Weinberg2023}, the helium yield required to achieve $Y = Y_{\odot}$ at $Z = Z_{\odot}$ is lower than any of our yield models predict. As a compromise intended to reproduce the solar helium abundance, our multi-zone models adopt $\yocc$ = 10$^{0.1} \zosun \approx 1.26 \zosun$ and the \nktext\ + V13 helium yields from Table \ref{tab:yield_interp}. We therefore adopt lower $\eta$ values to match the observed [O/H] gradient. With our adopted $\yocc$, Equation \ref{eta} gives $\eta$ = 0 at \rgal\ = 3.87 kpc, and we set $\eta$ to zero at smaller \rgal. The model's [O/H] gradient becomes flatter inside this radius, though not obviously inconsistent with the limited data available inside 5 kpc. In a forthcoming paper (J.Johnson et al., in prep.), we discuss the interplay of yields, outflows, star-formation history, and radial migration in determining gas and stellar metallicity gradients.

\citet{Johnson21} implement radial migration by matching stellar populations formed in {\tt VICE} to star particles from the {\tt h277} hydrodynamic cosmological simulation of \citet{Christensen2012}. This procedure becomes noisy for populations formed at large \rgal\ and early times, when the number of star particles available for matching is limited. L. Dubay et al. (submitted) implement an alternative scheme in which radial displacements are drawn from a Gaussian distribution whose dispersion as a function of birth radius and formation time is calibrated against the {\tt h277} results. We adopt this Gaussian migration scheme here, which makes our plots smoother but does not alter our conclusions regarding helium evolution relative to the original migration prescription. For time dependence of radial displacements, we adopt the ``diffusion" prescription of \citet{Johnson21}, with $\Delta R(t)$ = $\Delta R_{\mathrm{final}} \cdot \left(\Delta t / \Delta t_{\mathrm{final}}\right)^{1/2}$, where $\Delta t$ is the time elapsed since the population's birth and $\Delta R_{\mathrm{final}}$, the quantity drawn from the Gaussian distribution, is the displacement at the final time of the simulation, $t_{\mathrm{final}}$ = 12 Gyr.

As in Section \ref{sec: onezone}, we tie the metallicity dependence of the helium yield specifically to the oxygen abundance $\zo/ \zosun$. For results involving N and C, we adopt the N yields advocated by \citet{Johnson2023} and the C yields advocated by D. Boyea et al. (in prep), which are calibrated to reproduce observed trends for these elements with a combination of massive star and AGB yields. We discuss these yields further below.  For other details of the model, we refer the reader to \citet{Johnson21}.

\begin{figure}
\centering
    \includegraphics[width=0.9\columnwidth]{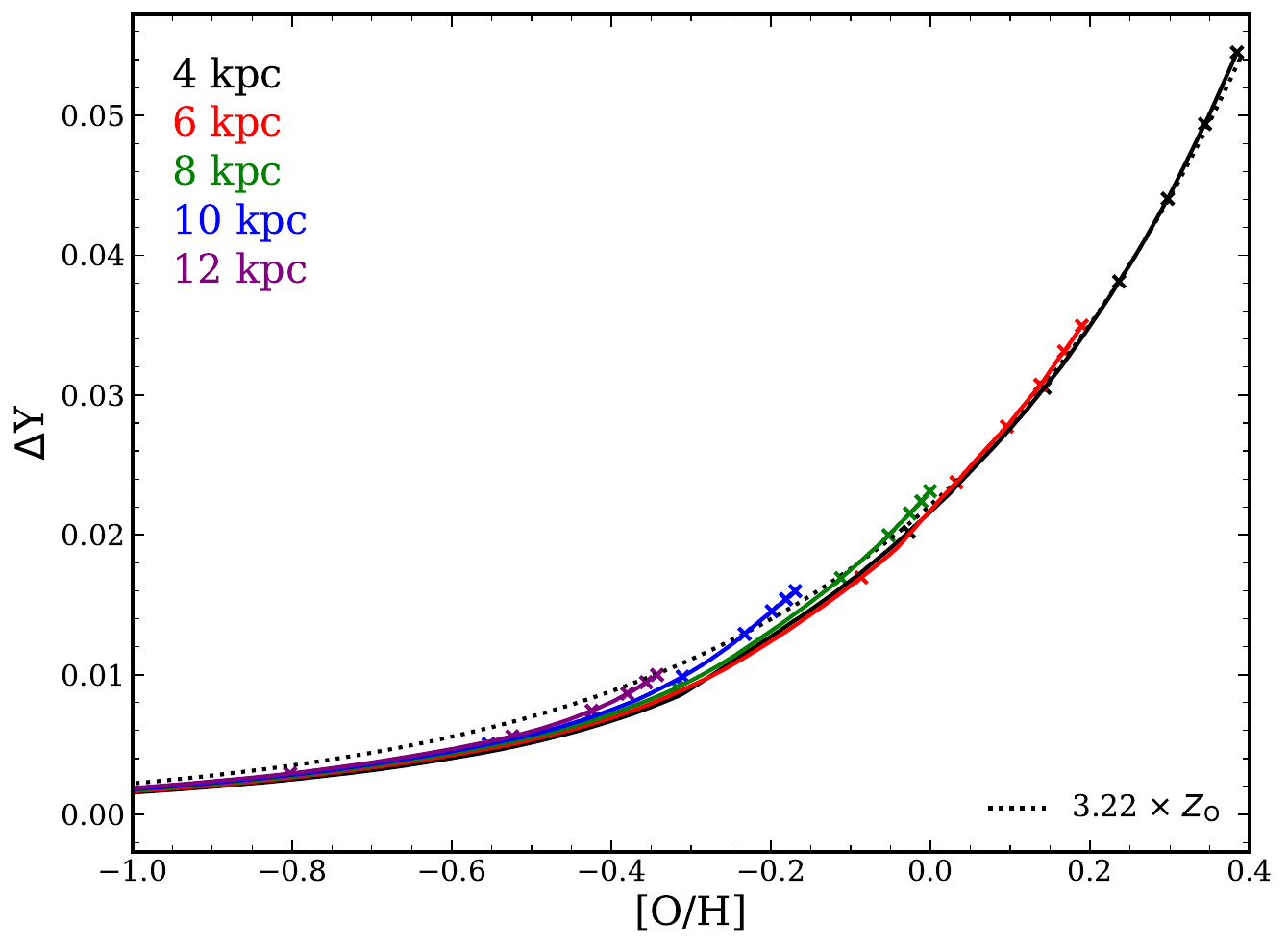}
    \caption{Gas phase $\dely$ vs [O/H] for our fiducial multizone model at fixed galactic radii (solid curves) over time. Points mark the abundances at times $t$ = 2, 4, 6, 8, 10, and 12 Gyr. The black dotted curve marks $\dely = 3.22 \zo$, with the slope corresponding to the yield ratio at solar [O/H].}
    \label{fig: yvsoh}
\end{figure}

\begin{figure*}
    \centering
    \centerline{\includegraphics[width=0.95\paperwidth]{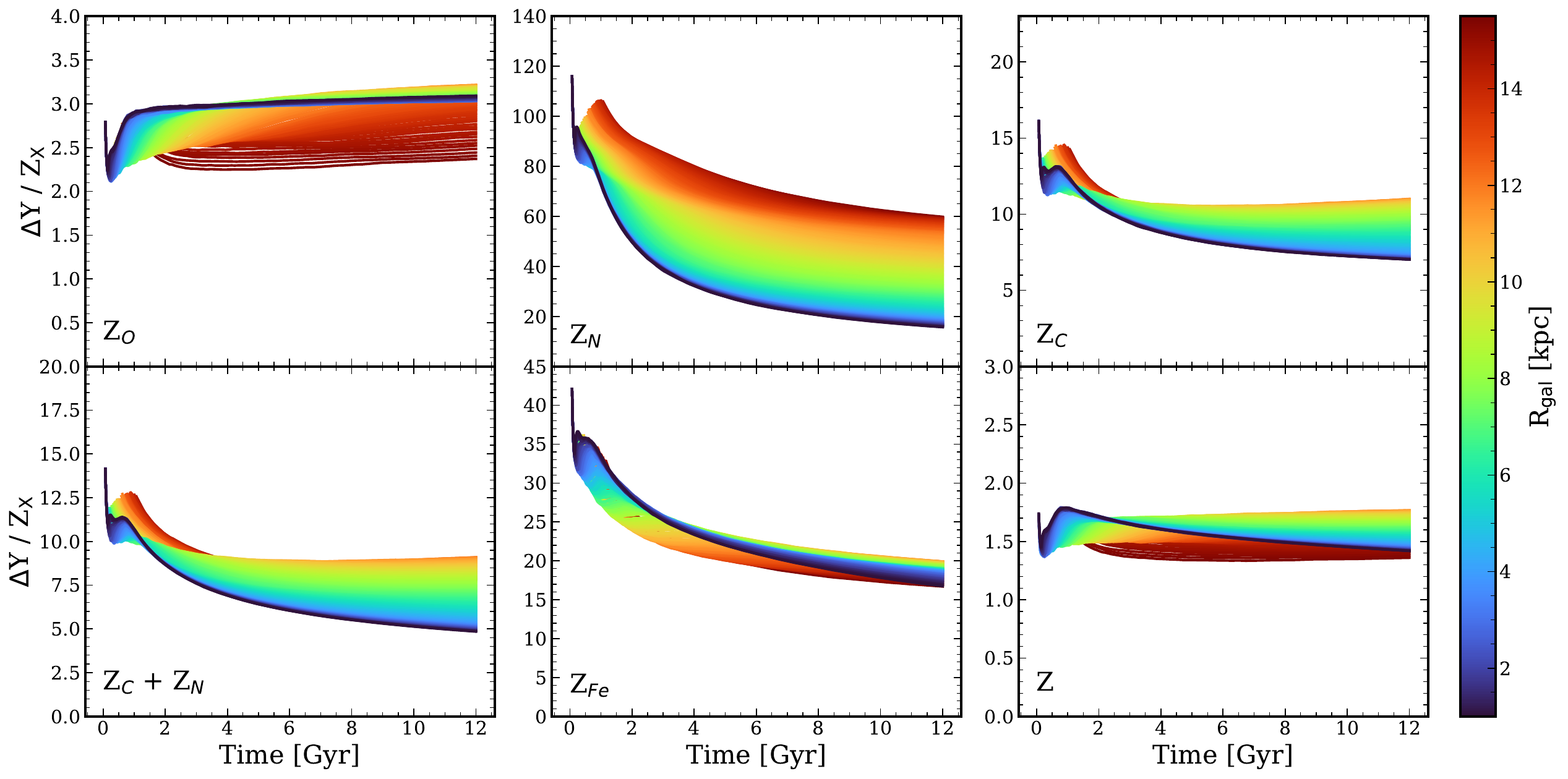}}
    \caption{The ratio of helium to different metals, denoted by $Z_{X}$, at different galactic radii over time. Curves are plotted beginning at [O/H] = -1.5 due to the sensitivity to the timestep of the numerical models at early times.}
    \label{fig: dydz}
\end{figure*}

\subsection{Results}
\label{sec: multizone_models}

Figure \ref{fig: yvsoh} plots the gas phase $\dely$ vs. [O/H] at radii \rgal\ = 4, 6, 8, 10, and 12 kpc. Points mark times $t$ = 2, 4, 6, 8, 10, and 12 Gyr. The simple appearance of this plot demonstrates a number of important points. First, the tracks at all radii are well described, though not perfectly described, by the relation $\dely \approx 3.22 \zo$, where $\zo = 0.0073 \cdot 10^{\mathrm{[O/H]}}$ is the gas phase oxygen mass fraction (dotted curve). Despite the combination of CCSN and AGB sources, mild metallicity dependence of the predicted $^4$He yields, and the impact of radially dependent star-formation history, outflows, and stellar radial migration, a linear $\dely - \zo$ relation remains an excellent approximation over the full range of time and \rgal\ (the lines appear curved because the $x$-axis is logarithmic). Second, tracks at smaller \rgal\ evolve to higher [O/H] and thus higher $\dely$ because of the lower $\eta$. Third, [O/H] evolution slows at late times as abundances approach equilibrium, with little change ($<$ 0.1 dex) over the final 4 Gyr. This slowing is more pronounced at large \rgal\ because the higher $\eta$ values lead to shorter evolutionary timescale $\bar \tau$ (Equation \ref{tbar}).

\begin{figure*}
    \centering
    \centerline{\includegraphics[width=0.95\paperwidth]{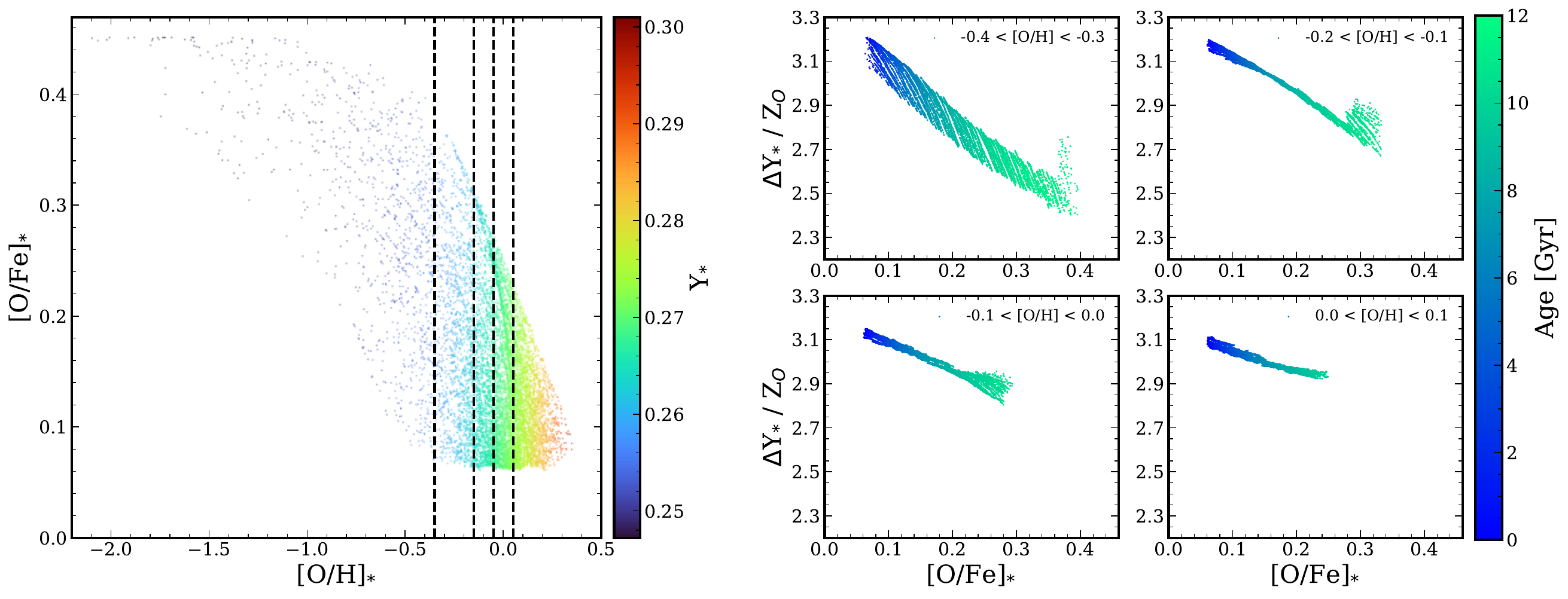}}
    \caption{\textit{Left:} [O/Fe] versus [O/H] for the solar annulus stars in our {\tt VICE} model color coded by the helium mass fraction. Dashed lines represent the chosen [O/H] values for binning. \textit{Right:} $\dely / \zo$ for the stars in the solar annulus as a function of [O/Fe], color coded by the age of the star and binned by the [O/H] value.}
    \label{fig: ofe_oh}
\end{figure*}

\begin{figure*}
    \centering
    \centerline{\includegraphics[width=0.95\paperwidth]{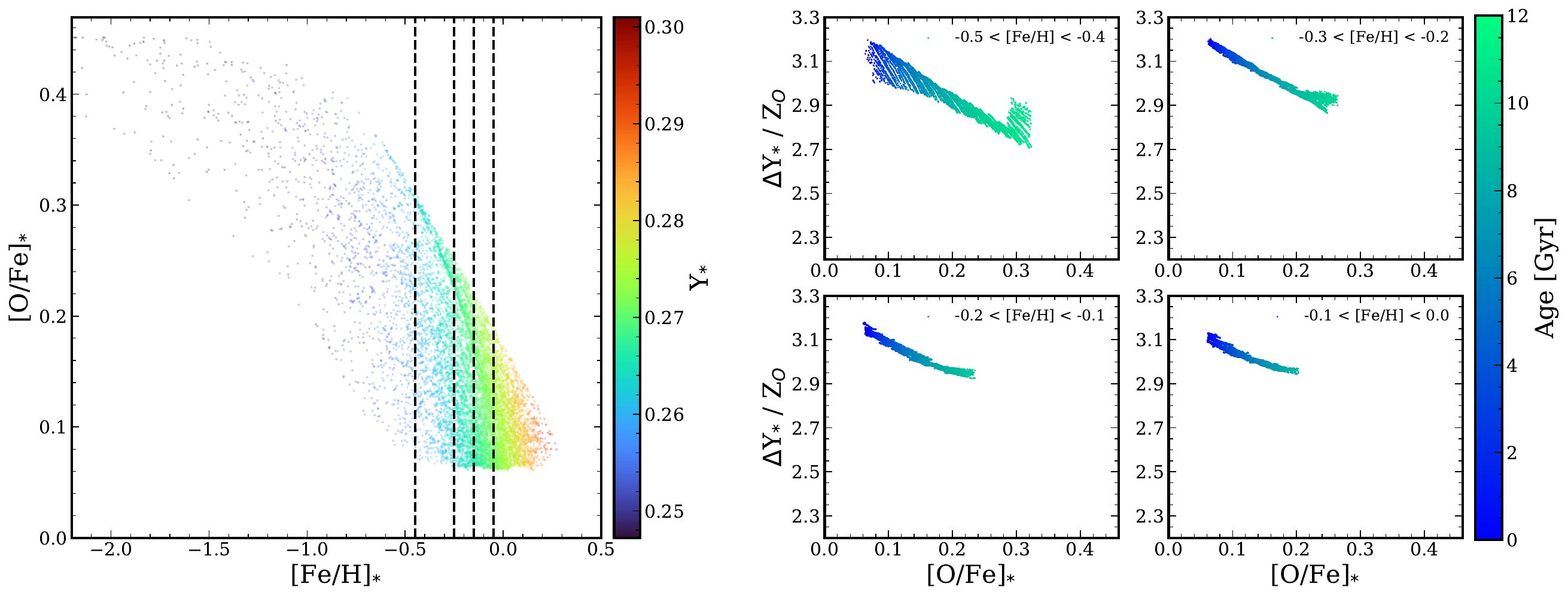}}
    \caption{Same as in Figure \ref{fig: ofe_oh}, but using [Fe/H]. The right panels still plot $\dely_{*} / \zo$, but with bins of [Fe/H] rather than [O/H].}
    \label{fig: ofe_feh}
\end{figure*}

Figure \ref{fig: dydz} examines the gas phase $\dely / Z$ relation in greater detail, for several elements, plotting $\dely / Z_{X}$ vs. time for all zones from 1 -- 15.5 kpc. Numerical artifacts can arise at very early times because of the finite timestep and the large ratio of accreted helium to synthesized helium. We mitigate these artifacts by only plotting curves once [O/H] > -1.5, but small bumps in the first Gyr should still be considered unreliable.

At all times, $\Delta Y / Z_{\rm O}$ is lower in the outer Galaxy because the metallicity is lower and the predicted massive star yield is lower (Figure \ref{fig: imf_avg}, Table \ref{tab:yield_interp}). At a given radius, $\Delta Y / Z_{\rm O}$ increases slightly with time because of growing metallicity and the delayed contribution of AGB stars. We caution that the predicted spread between the inner and outer Galaxy depends sensitively on the metallicity dependence of the IMF-averaged yields. For $R < 8\ \mathrm{kpc}$, where [O/H] > 0 at late times, this metallicity dependence relies on extrapolation of the LC18 yields beyond the highest metallicity models at [M/H] = 0 (see Figure \ref{fig: imf_avg}).\footnote{\cite{Roberti2024} recently published yields at $Z=2Z_\odot$, which we will use in future work.} This is true even for the \nktext\ yields adopted here because we use LC18 to compute the contribution of stars with $m$ > 40 \M. The most robust conclusion is that the predicted $\Delta Y / Z_{\rm O}$ is nearly constant, with a small dependence on time and \rgal.   The overall level depends on $(\yheagb + \yhecc) / \yocc$, which we have chosen to give solar $\Delta Y$ at solar $\zo$. We have assumed that $\yocc$ itself is metallicity-independent, and variations in $\yocc$ with metallicity would affect the appearance of this plot.

Figure \ref{fig: dydz} also shows $\dely / \zfe$ vs time. Here, there is significant evolution of the ratio, dropping by roughly a factor of two, but this evolution is driven by the addition of SNIa Fe to $\zfe$, not by $\dely$. We have implemented metallicity dependence of helium yields by tying it to $\zo$, and if we instead tied it to $\zfe$, then both the $\dely - \zo$ and $\dely - \zfe$ trends would change. However, the metallicity dependence of the helium yield is a small effect compared to SNIa Fe enrichment, so we would still expect a nearly flat trend for $\dely - \zo$ and a time-dependent trend for $\dely - \zfe$.

For $\dely - \zn$, the behavior is strikingly different from $\dely - \zo$, mainly driven by the metallicity dependence of the N yields. Based on the arguments of \citet{Johnson2023}, which draw on theoretical yield models and empirical constraints, we take $\yncc = 0.024 \cdot \yocc$, independent of metallicity, and the AGB yields predicted by C11 + C15 scaled by a factor of $\left(\yocc / 0.005\right)=1.84$. The IMF-averaged $\ynagb$ is approximately linear in $Z$ and equal to $\yncc$ at $Z \approx 0.0387\ Z_{\odot}$. Our overall yield normalization gives $\zn \approx Z_{\mathrm{N}, \odot}$ at $Z = Z_{\odot}$. The metallicity dependence of $\yncc + \ynagb$ reproduces observed [N/O] vs. [O/H] trends in the MW and external galaxies \citep{Johnson2023}. The predicted $\dely / \zn$ is higher at large \rgal\ because of the lower N yield. At a given \rgal, $\dely / \zn$ declines with time mainly because increasing metallicity leads to higher N yield, and secondarily because the relative contribution of AGB stars increases with time. Because AGB N comes largely from stars with $m$ > 2 \M, the time-delay of AGB enrichment is quite short \citep{Johnson2023}, similar to our findings here for helium.

For $\dely / \zc$, results are qualitatively similar to those for $\dely / \zn$, but the trends are weaker because the metallicity dependence of C yields is weaker, both in theoretical production and from empirical trends (D. Boyea et al., in prep.). Here, the massive star C yield increases linearly with metalliciy, $\yccc = 5.28\times10^{-2} \left(Z - 0.014\right)\ +\ 2.28\times10^{-3}$, while the adopted AGB yield is $1.36\times$ the C11 + C15 table. In surveys of evolved stars, the number-weighted C + N abundance is often used in place of C or N individually because this C + N combination is conserved when CNO-processed material is mixed into the photosphere by dredge-up (see, e.g., \citet{Shetrone2019, Roberts2024}). Ignoring the small impact of the C and N atomic mass difference, here we simply divide by $(\zc + \zn)$. The overall behavior of $\dely / (\zc + \zn)$ is similar to that of $\dely / \zc$, which is unsurprising as C dominates the abundance.

Overall we favor the use of $\alpha$-elements such as O, Mg, or Si as a reference for $\dely$, since helium nucleosynthesis is dominated by massive stars and AGB stars with a relatively short time delay. While N and C have a similar combination of enrichment sources, it is primarily the metallicity dependence of yields that drives variation in $\dely / Z_{\mathrm{X}}$, and that metallicity dependence is less well understood for these elements. The final panel of Figure \ref{fig: dydz} plots $\dely / Z$ where $Z$ is the overall metal mass fraction. Because {\tt VICE} does not track all elements, we compute $Z$ as
\begin{equation}
    \label{ztot}
    Z = Z_{\odot}  \left({\sum Z_{i} \over \sum Z_{i, \odot}}\right)~,
\end{equation}
where the sum is over all elements heavier than helium tracked in the simulation, which in our case corresponds to O, C, N, and Fe, and we take $Z_{\odot}$ = 0.014. We predict a flat age trend and relatively narrow spread in $\dely / Z$. Our derived ratio is $\dely / Z \approx \dely_{\odot} / Z_{\odot} = 1.64$, which is not surprising because we have calibrated our yields to produce solar abundances. 

Figure \ref{fig: ofe_oh} shows helium abundances of stars rather than gas, specifically stars that have 7 kpc < \rgal\ < 9 kpc and |$z$| < 2 kpc at the present day. The left panel plots [O/Fe] vs. [O/H] with points color-coded by $Y$. Although the distribution of [O/Fe] is not bimodal, there is a range of [O/Fe] at each value of [O/H] because different Galactic zones follow different [O/Fe]-[O/H] tracks as a result of differing $\tstar$ and $\eta$. The stars with the highest [O/H] at a given [O/Fe] have migrated to the solar annulus from the inner Galaxy where the equilibrium abundance is high, and low [O/H] stars have migrated from the outer Galaxy. More metal-rich stars have higher helium abundance as expected. However, there is a slight trend of $Y$ with [O/Fe] at fixed [O/H], which is shown more clearly on the right hand panels where we plot $\dely / \zo$ vs. [O/Fe] for stars with a narrow range of [O/H]. The maximum value of $\dely / \zo$ changes from panel-to-panel because of the metallicity dependence of the helium yield, but within a given panel for fixed [O/H], that effect is removed. Instead, $\dely / \zo$ rises towards lower [O/Fe] because the lower [O/Fe] stars are younger, leaving more time for helium enrichment from lower mass AGB stars. We have confirmed this explanation by running models with instantaneous AGB enrichment but the same IMF-averaged yield, in which case the trend of $\dely / \zo$ vs. [O/Fe] is flat within each [O/H] bin. Nonetheless, this sloping effect is most noticeable for the -0.5 < [O/H] < -0.4 bin because of the larger range of [O/Fe], with $\dely / \zo$ varying from 2.4 to 3.2. This prediction might be testable with asteroseismic $Y$ measurements, though it requires the ability to confidently distinguish 20\% differences in $\dely$, or roughly 0.002 in $Y$ at these metallicities.

Figure \ref{fig: ofe_feh} resembles Figure \ref{fig: ofe_oh}, but now we use [Fe/H] on the $x$-axis of the left panel and take narrow bins of [Fe/H] in the right panels. The trends with [O/Fe] at fixed [Fe/H] are flatter than those at fixed [O/H], because at fixed [Fe/H], the stars with higher [O/Fe] have higher $\zo$ and higher overall metallicity. The metallicity dependence of the helium yield partly cancels the impact of greater age at larger [O/Fe]. The relative appearance of Figures \ref{fig: ofe_oh} and \ref{fig: ofe_feh} would be different if we tied the helium yield to $\zfe$ instead of $\zo$.

The findings from our multi-zone model are noticeably different from those of the simulation-based study of \cite{Vincenzo2019} in several respects (see, e.g., their Figure 5).  \cite{Vincenzo2019} find significant scatter in the gas-phase $Y-\zo$ relation and $\sim 15\%$ differences in the slope of this relation among their three simulated galaxies.  They find less scatter in the trends between $Y$ and $\zc$, $\zn$, and $\zc+\zn$, and they therefore advocate the use of these elements as references for helium evolution.  We are unsure of the reasons for these differences, though several factors could contribute.  The \cite{Vincenzo2019} simulations use the K10 AGB yields and the NKT13 massive star yields; because the latter are lower than the NKT13$_{\rm EXT}$ or LC18 yields that we adopt, the relative contribution of AGB helium enrichment is higher.  These differences in helium yields would influence the $\Delta Y/Z_{\rm X}$ slopes, but it is not obvious why they would introduce scatter in $\Delta Y/\zo$.  Differences might also arise from our choice of O, C, and N yields; if the \cite{Vincenzo2019} O yield has a strong metallicity dependence, then the $\Delta Y/\zo$ ratio would change with metallicity and enrichment history. Cosmological simulations include physical effects that are not represented in our more idealized model.  It is again not obvious how these effects would introduce $\Delta Y/\zo$ scatter, but one possibility is that feedback-induced outflows have different impact on CCSN and AGB elements, inducing scatter in the CCSN/AGB enrichment ratio at fixed metallicity.  Finally, we note that the mass of gas particles in the \cite{Vincenzo2019} simulations is about $10^7 M_\odot$, so the star-forming components of the forming galaxies are resolved by $10^3-10^4$ particles.  Discreteness effects might induce scatter in abundance ratios, and they might affect CCSN and AGB enrichment differently because of their different timescales.

\subsection{Comparisons to Observations}
\label{sec: multizone_obs}

As noted in the introduction, helium abundance measurments are challenging both in gas phase and in stars. Nonetheless, there are measurements at a range of metallicities from HII regions, from star clusters, from hot stars, from RR Lyrae, and, of course, from the Sun. Asteroseismic measurements may enable future helium studies for much larger stellar samples. Where a study reports only the number ratio of helium, He/H = $y = n_{\mathrm{He}} / n_{\mathrm{H}}$, we transform to a mass fraction with the widely used equation, originally from \citet{Peimbert1974, Peimbert1976}
\begin{equation}
    \label{he_num_to_mass_oxy}
    Y = \frac{4y (1 - 20 * \mathrm{O/H})}{(1 + 4y)}.
\end{equation}
The factor of four in the numerator converts $^4$He number to mass, the (1 + 4$y$) denominator accounts for helium contribution to the total mass, and the $(1 - 20 n_{\mathrm{O}} / n_{\mathrm{H}})$ factor approximately accounts for the impact of metallicity on $n_{\mathrm{H}}$.

\begin{figure}
\centering
    \includegraphics[width=0.9\columnwidth]{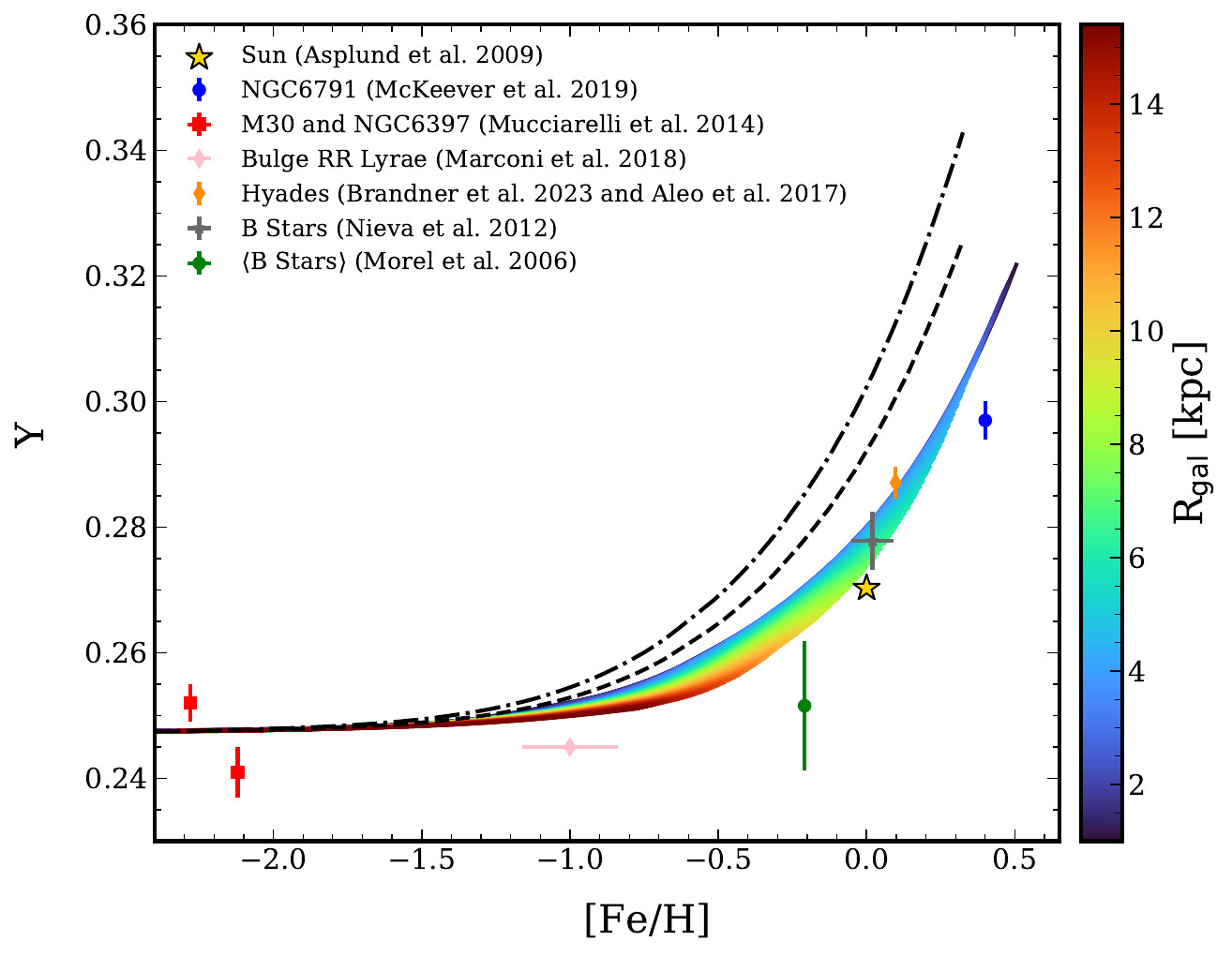}
    \caption{Helium gas mass fraction as a function of gas phase [Fe/H], compared to stellar measurements. The rainbow curves are the \nktext\ + V13 {\tt VICE} models while the dashed and dot-dashed black curves are the LC18 + V13 and LC18v300 + V13 models, respectively. Red points are the M30 and NGC6397 globular clusters \citep{Mucciarelli14}, the pink point represents Milky Way Bulge RR Lyrae stars \citep{Marconi18}, the green and gray points represent the mean value for B stars in the solar neighborhood from \citet{Morel2006} and \citet{Nieva12},  the orange point and blue points are the Hyades \citep{Aleo_2017, Brandner2023} and NGC6791 \citep{McKeever19} open clusters. The initial abundance of the Sun is represented by the yellow star from \citet{Asplund2009}.}
    \label{fig: star_obs}
\end{figure}

Figure \ref{fig: star_obs} plots the helium mass fraction as a function of [Fe/H]. The rainbow of curves plots the evolution of gas phase abundances in each radial zone for our fiducial model. Every star in the simulation was born from gas lying along one of these curves, so the rainbow shows the maximum spread in stellar $Y$ predicted at any [Fe/H]. For \rgal\ = 4 kpc, we also show the predicted tracks if we adopt the LC18 + V13 helium yields (dashed curve) or the LC18v300 + V13 helium yields (dot-dashed curve) while keeping other yield and model parameters fixed. Colored points show $Y$ estimates from two globular clusters at $\feh<-2$, from Bulge RR Lyrae stars at $\feh=-1$, from two studies of B stars at near solar metallicity, from the \citet{Asplund2009} solar abundance, from the Hyades at $\feh \approx +0.14$, and from the old, massive open cluster NGC6791 at $\feh\approx +0.4$. \citet{Marconi18} report that bulge RR Lyrae are ``consistent with the primordial helium abundance" with a value of $Y$ = 0.245 without assigning a vertical error bar. The fiducial model matches the solar abundance essentially by construction (through our choice of yield normalizations), and having done so, it is consistent with the remaining stellar data except for the \citet{Morel2006} B-star point, which gives a surprisingly low $Y \approx \yp$ for slightly sub-solar [Fe/H], although it is close to being within the uncertainty. In detail, the model slightly underpredicts the estimated $Y$ of the Hyades and overpredicts the estimated $Y$ of NGC6791, which is an old and massive open cluster.

If we adopt LC18v0 or LV18v300 massive star yields in place of \nktext, the model overpredicts all of the $Y$ estimates at [Fe/H] $\geq 0$, including the Sun's. This finding is expected given our results for one-zone models in Section \ref{sec: onezone}, and Equation \ref{he_yield_scaled} in particular. The caveat is that we could accommodate higher helium yields if we also raised $\yocc$ and increased outflows to match the MW [O/H] gradient.

\begin{figure}
\centering
    \includegraphics[width=0.9\columnwidth]{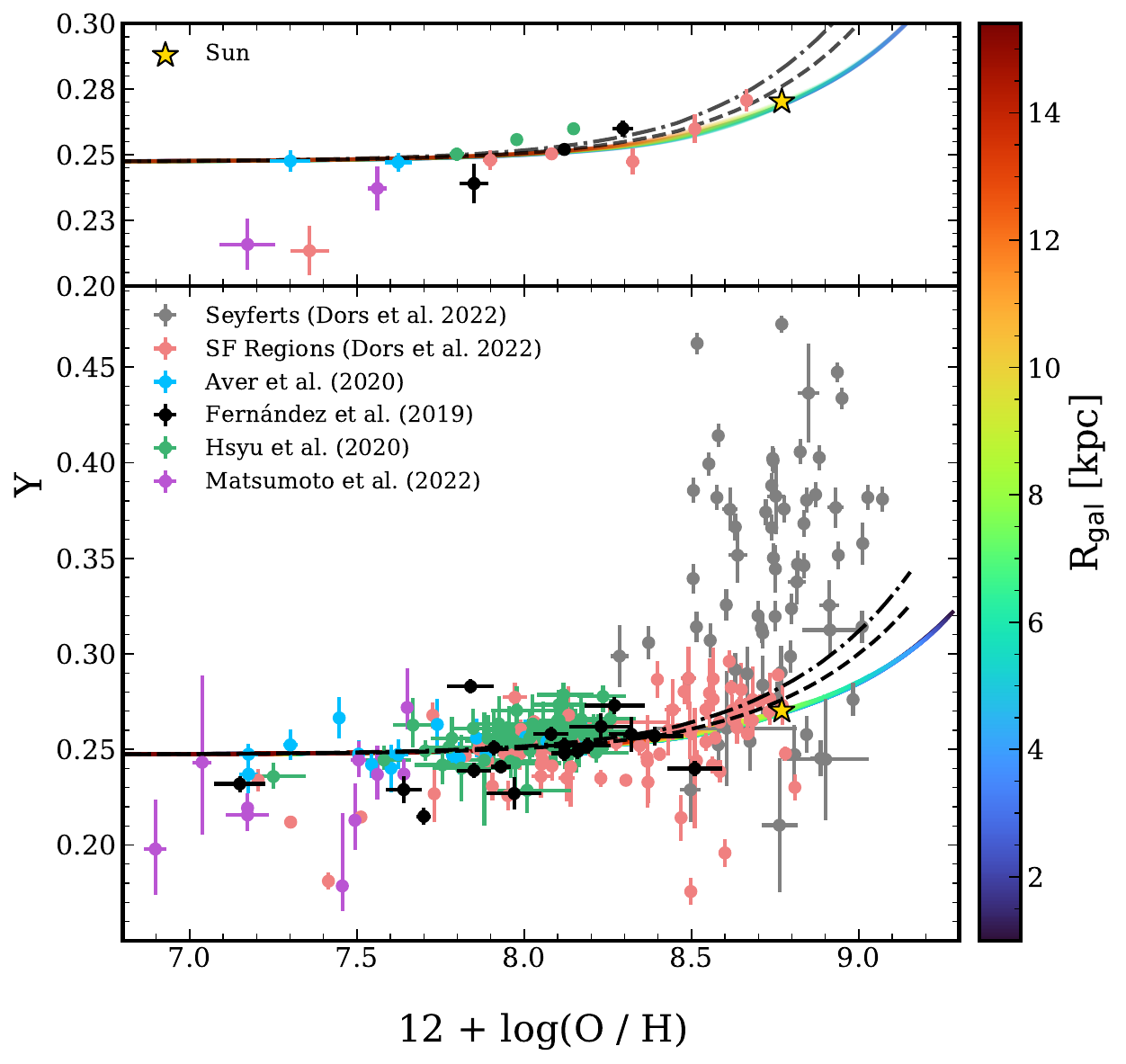}
    \caption{Helium gas mass fraction as a function of gas phase oxygen abundance. The rainbow curves are the \nktext\ + V13 {\tt VICE} models while the dashed and dot-dashed black curves are the LC18 + V13 and LC18v300 + V13 models, respectively. Blue points are from \citet{Aver2021}, black points are from \citet{Fernandez2019}, pink and gray points are from \citet{Dors2022}, purple points are the 10 Subaru galaxies from \citet{Matsumoto2022}, and the green points are from Sample 1 of \citet{Hsyu2020}. The upper panel plots median values in bins of [O/H], requiring a minimum of four measurements per bin.}
    \label{fig: gas_obs}
\end{figure}

Figure \ref{fig: gas_obs} compares our models to gas-phase $Y$ measurements from external galaxies. Many studies target low metallicity galaxies to constrain the primordial helium abundance \citep[e.g.,][]{Fernandez2019, Aver2021, Hsyu2020, Matsumoto2022}. \citet{Dors2022} estimate $Y$ in Seyfert 2 galaxies and in a compilation of extragalactic star forming (SF) regions. Comparing to the full set of individual data points (lower panel), the impression is of reasonable agreement between the model and data, except for the Seyfert 2 determinations, which appear anomalously high. However, many individual measurements are inconsistent with others at similar (O/H), at a level that is significant relative to the reported uncertainties, probably a reflection of systematic errors that are not captured by these statistical error bars. In the upper panel, we plot the median $Y$ values in bins of 12 + log(O/H) = 7.0 -- 7.5, 7.5 -- 8.0, and 0.2-dex wide at higher (O/H). We only plot points for a given study and [O/H] bin if it includes at least four systems, and we assign an error bar that is the dispersion of the measurements divided by $\sqrt{N}$, where $N$ is the number of systems in the bin. In this representation, we see good agreement with \citet{Aver2021} and \citet{Hsyu2020} and some but not all of the \citet{Fernandez2019} and \citet{Dors2022} SF points. Both \citet{Matsumoto2022} points are below our predictions, but these have rather high uncertainties. Other than the Seyfert 2 data, there are few extragalactic gas phase measurements at high enough metallicity to map out the expected rise above the primordial abundance. 

\begin{figure}
\centering
    \includegraphics[width=0.9\columnwidth]{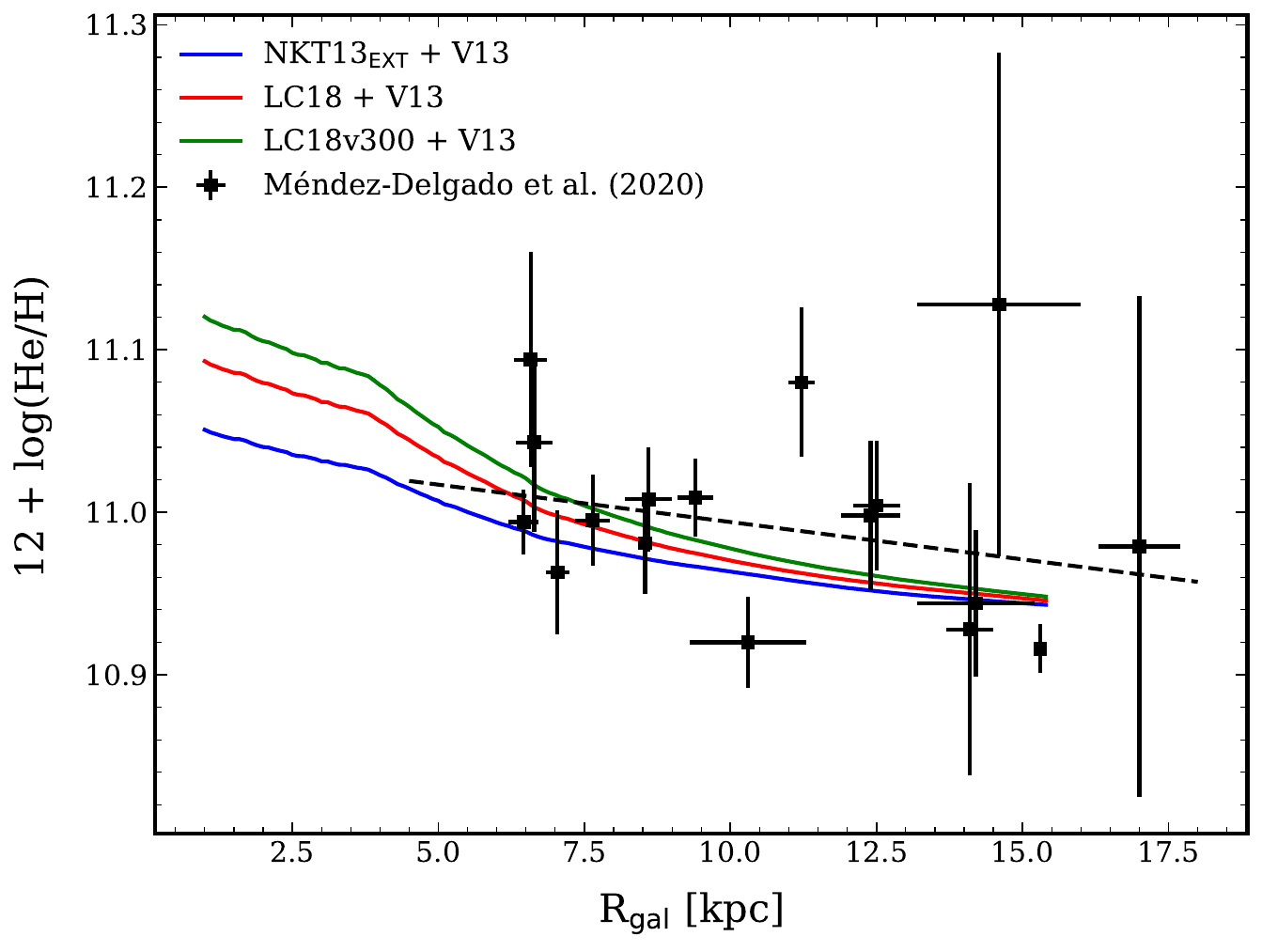}
    \caption{Radial gradient of the helium abundance in our simulations and in the Milky Way. The green, red, and blue curves represent the LC18v300 + V13, LC18 + V13, and \nktext\ {\tt VICE} models, respectively. The black points are from \citet{Delgado2020} where we only take values from the HII regions in their sample. The black dashed curve is their best fit to the data.}
    \label{fig: rad_grad}
\end{figure}

Figure \ref{fig: rad_grad} shows the radial gradient of helium from HII regions in the Milky Way from \citet{Delgado2020} in the black points. The black dashed line is their best fit to the data. The different colored curves are our different {\tt VICE} models. Due to the lack of readily available oxygen data, we have to use different units and cannot apply Equation \ref{he_num_to_mass_oxy}. As such, to construct the gradient, we take the last reported [He/H] value in any radial zone and transform it to the logarithmic scale where hydrogen is typically 12, assuming that Solar $Y$ is 0.2703 and Solar $Z$ is 0.014. Thus, He/H in the Sun is approximately 0.0944 and 12 + log(He/H) is approximately 10.97. Given the large statistical uncertainties and scatter in the data points, it is difficult to draw strong conclusions, but the mild gradient predicted by our models is not obviously inconsistent with MW observations.

\begin{figure}
\centering
    \includegraphics[width=0.9\columnwidth]{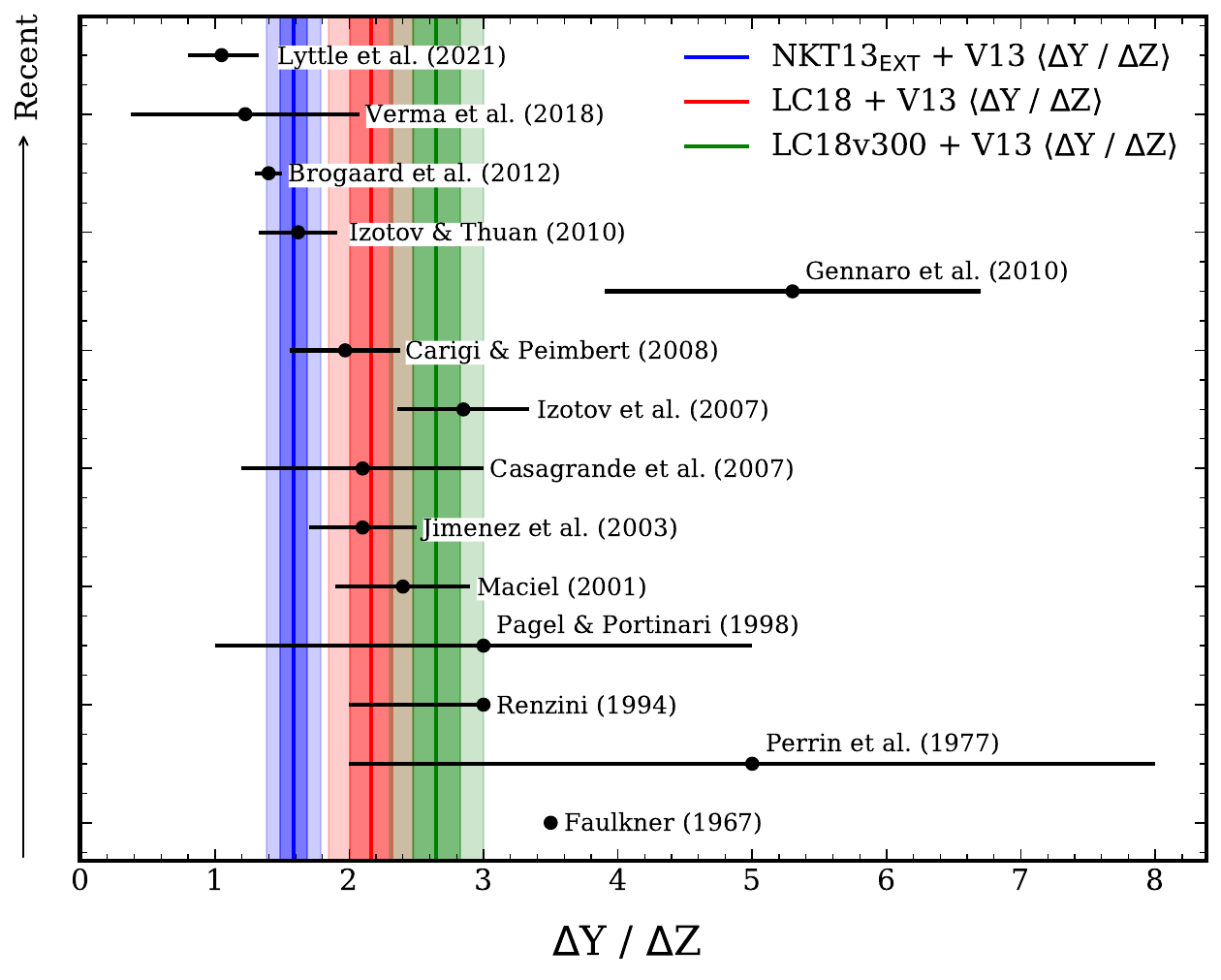}
    \caption{List of literature $\dely / \Delta Z$ values spanning several decades. In more recent times (i.e., \citet{Lyttle2021, Verma2019, Brogaard2012, Izotov2010, Carigi2008, Izotov2007, Casagrande2007} with the exception of \citet{Gennaro2010}), the uncertainties have decreased and have approached values closer to 1. The earliest measurements (i.e., \citet{Faulkner1967, Perrin1977, Renzini1994, Pagel1998, Maciel2001, Jimenez2003}) tend to have greater error bars and higher values. The green, red, and blue lines represent the mean values for the LC18v300 + V13, LC18 + V13, and \nktext\ + V13 {\tt VICE} models, respectively, while the shaded regions correspond to the 1$\sigma$ and 2$\sigma$ ranges. To calculate the average values from the models, we take the final reported $\dely / \Delta Z$ value in a zone, where $Z$ is the scaled-solar oxygen abundance, and then take the average of all of the radial zones.}
    \label{fig: lit_dydz}
\end{figure}

Finally, Figure \ref{fig: lit_dydz} compares the average value of $\dely / \Delta Z$ from this study to a comprehensive list of values available in the literature. For this plot, $Z$ represents our model calculation of the scaled-solar oxygen mass abundance instead of the scaled-solar addition of all of the elements tracked in the simulation that was used in the bottom right panel of Figure \ref{fig: dydz}. Over time, the observational uncertainty on this ratio has decreased, and seems to be trending to a value between 1 and 2. Our fiducial model predicts a value of around 1.6, which is consistent with most recent studies. The predicted values depend mainly on the yield ratio $\left(\yhecc + \yheagb\right) / \yocc$ along with our adopted values of $\zosun$ and $Z_{\odot}$.

As discussed in Section~\ref{sec: multizone_models}, the results of the \cite{Vincenzo2019} simulations differ from those of our multi-zone model in predicting larger scatter between $\Delta Y$ and $\zo$ and a tighter correlation of $Y$ with C and N.  Current observations are not sufficient to distinguish these predictions, but doing so would be interesting, especially if the scatter in these relations can be mapped over a range of metallicity and age.  If there is large scatter in $Y$ at fixed $\zo$ (or for other $\alpha$-elements $Z_{\alpha}$), it would suggest that some mechanism can cause large fluctuations in the ratio of AGB to CCSN enrichment at fixed metallicity, or some more basic flaws in our current understanding of helium yields.  Similarly, it would be valuable to test whether the $\dely/\zo$ ratios are similar in low metallicity galaxies and the Sun, as our models predict (Figure~\ref{fig: he_lowmet}).  A different ratio at low metallicity would imply a stronger than expected metallicity dependence of helium and/or oxygen yields, and additional elements could be used to assess the relative importance of the numerator and denominator in driving this metallicity dependence.

%% file: 5Conclusions.tex
\section{Summary}
\label{sec: conclusions}

We have investigated the galactic chemical evolution of $^4$He using both one-zone and multi-zone GCE models, with particular attention to yields and empirical constraints. Although helium is the most abundant product of stellar nucleosynthesis, the yield from massive stars depends on how much of the synthesized helium is ejected rather than converted into C and O. Much of the massive star yield is released in pre-supernova winds rather than the terminating explosions, and particular care is required in computing \textit{net} yields because a star's birth helium is often large compared to the newly produced helium. The IMF-averaged yield differs by a factor of 2 -- 3 between NKT13 and LC18, two widely used massive star studies. Some of this difference arises because NKT13 do not include yields from stars with birth mass $m$ > 40 \M. We therefore defined \nktext, which combines NKT13 yields for stars with 8\M\ $\leq m \leq$ 40 \M\ and LC18 yields for stars with 40\M\ $\leq m \leq$ 120 \M. We find reasonable agreement in the IMF-averaged yield of NKT13 and LC18 at low metallicity, but at solar metallicity the NKT13 yield is about $1.5\times$ lower and agrees well with the solar metallicity prediction from S16. All of these yields are calculated for non-rotating stellar models, but helium synthesis can be strongly affected by rotational mixing; the $v = 300$ km/s models of LC18 predicted an IMF-averaged yield 30-40\% higher than their non-rotating models.

In all of the massive star yield models we consider, stars with $m$ = 40 -- 120 \M\ contribute 40 -- 60\% of the IMF-averaged helium yield (except for NKT13, which does not include such stars). The predicted yield is therefore sensitive to the assumed IMF slope and maximum mass, for which we adopt \citet{Kroupa} and 120 \M, and to the models of these high mass stars. Except for NKT13, these studies do not go to $Z > Z_{\odot}$, so we have little theoretical guidance on the metallicity dependence of helium yields in the super-solar regime.  The recently published yields of \cite{Roberti2024} will remedy this situation.

In contrast to massive stars, the predicted IMF-averaged helium yields from intermediate mass AGB stars appear fairly robust. We find good agreement between V13 and C11 + C15, and somewhat higher ($\sim$ 30\%) yields from K10. The yields from KL16 + K18 are substantially different, probably because of issues with the stellar evolution calculations discussed by \citet{Pons2022}. In the other three models, about half of the AGB yield comes from stars with $m >$ 4.5 \M, so the characteristic time for a stellar population to release most of its AGB helium is quite short. In all three models, we find that half of the IMF-averaged helium yield is released in $\lesssim$ 300 Myr, and 75-80\% in 1 Gyr. This makes AGB helium enrichment fast compared to SNIa Fe enrichment, where roughly half is released in 1 Gyr for a conventional delay time distribution.

In comparison to the \nktext\ massive star yield, the V13 AGB yield is similar in magnitude at low $Z$ and about a factor of two lower at solar metallicity. The sum of the two yields is only mildly dependent on metallicity (Table \ref{tab:yield_interp}). For LC18, the massive star yields are larger, and the metallicity dependence is stronger. We again caution that our results for $Z > Z_{\odot}$ rely largely on extrapolation. 

Although helium enrichment includes prompt CCSN and delayed AGB contributions, each of them metallicity dependent, we find that results from numerical one-zone models that incorporate all of these effects are well approximated by an analytic model, Equations \ref{he_analytic} and \ref{yeq}, that ignores metallicity dependence and treats AGB yields as instantaneous. This accuracy arises because AGB enrichment is still fairly prompt and metallicity dependence of the combined contribution is mild. If the IMF-averaged oxygen yield is metallicity-independent, as we assume, then the analytical model simply predicts $\dely / \zo  = \left(\yhecc + \yheagb\right) / \yocc$. One can infer this yield ratio from solar $\dely$ and $\zo$ (Equation \ref{he_yield_scaled}). Although there is a transition from CCSN to CCSN + AGB enrichment, we find that even the low metallicity galaxies studied for $\yp$ should be well described by a linear $\dely - \zo$ relation, and for the yield models studied here, the slope in the low metallicity regime should still be close to $\dely_{\odot} / Z_{\odot}$.

For our fiducial $\delysun$ = 0.023 \citep{Asplund2009} and $\yocc = 0.973 \times \zosun$ = 0.0071 \citep{Weinberg2023}, we infer $y_{\mathrm{He}}$ = 0.022, i.e., for every solar mass of stars formed, massive stars and AGB stars return an average of 0.022 \M\ of new helium, in addition to about 0.08 \M\ $(= 0.25r)$ returned by recycling of birth helium. This net yield is lower than predicted by the combination of V13 AGB yields and our lowest yield massive star model \nktext, which gives $\yheagb + \yhecc = 0.029$ at solar metallicity. For the multi-zone GCE models of Section \ref{sec: multizone}, therefore, we bump our $\yocc$ value upwards by 0.1 dex relative to the central value of \citet{Weinberg2023}, and we adopt \nktext\ and V13 for massive star and AGB yields, respectively.

The multi-zone {\tt VICE} models include radially dependent star-formation history, star-formation efficiency, and outflow efficiency, and they incorporate time-delayed AGB enrichment, continuous recycling, and stellar radial migration. Nonetheless, the end results are quite simple. To a good approximation, the gas-phase abundance evolves along a track of constant $\dely / \zo$, with a rate and endpoint that changes with \rgal. Stars inherit this near-constant $\dely / \zo$ from the gas. In detail, we find that higher [O/Fe] stars have slightly lower $\dely / \zo$ at fixed [O/H] because older stars have less time for AGB enrichment. This prediction might be testable with asteroseismic $Y$ measurements. As a reference element for $\dely / Z_{\mathrm{X}}$ ratios, we recommend an $\alpha$-element with a yield that is expected to be nearly metallicity independent, such as O, Mg, or Si. Because the predicted AGB enrichment is fast compared to SNIa enrichment, helium tracks these elements more closely than it tracks Fe, to the extent there is a difference. The expected but uncertain metallicity dependence of C and N yields complicates their use as $\dely$ tracers.

With our adopted yields, our model predictions appear consistent with most observations of stellar and gas-phase helium abundances in the Milky Way and external galaxies, with the high $Y$ values inferred for Seyfert 2 galaxies by \citet{Dors2022} as an exception. With higher ratios of $y_{\mathrm{He}} / \yocc$, we would be unable to reproduce the observed [O/H] gradient and the observed Milky Way trends simultaneously, and we would not expect to produce solar $\dely$ at solar metallicity. We do not expect our predictions to apply to globular clusters that self-enrich with substantially different yields, e.g. because they retain the nucleosynthetic products of some of their stars but not others.

While the stellar astrophysics of helium nucleosynthsis is complex, our models suggest a simple picture for galactic evolution of helium given the IMF-averaged massive star and AGB yields. The model predictions can be more stringently tested by measurements in HII regions at solar and super-solar metallicity, and by further open cluster measurements and asteroseismic measurements that probe stars with a range of age, [$\alpha$/Fe], and environment, as well as metallicity. These measurements could confirm the simple picture implied by our models, or they could show complexities that reveal surprises in the stellar or galactic astrophysics of the second most abundant element.

%% file: 6Appendix.tex
\section{Yields}
\label{appendix: yields}

We provide three files available for download\footnote{https://github.com/miqaela708/Weller24} to implement the yields from this paper into {\tt VICE}, as we are currently working on building an empirical yield module for future release. These are 1) NKT13\_Selection.txt, 2) LC18\_Selection.txt, and 3) yield\_settings.py. The only file that is modifiable is yield\_settings.py, and modifications should be confined to the parameter choices specified in the first $\sim 20$ lines. All of the default yield sets used in this paper are already selected, while alternative allowed yield sets are shown next to the appropriate variable. 

Instructions on installing {\tt VICE} can be found on PyPI,\footnote{https://pypi.org/project/vice/}.  To run with our yields, these three additional files need to be imported under the directory {\tt VICE/migration/src/simulations/}, and the {\tt disks.py} file will need the following lines removed or commented out as they are default yield presets: \\ \\
{\tt from vice.yields.presets import JW20 \\ vice.yields.sneia.settings['fe'] *= 10**0.1} \\ \\
After these lines are commented out, import the following line in {\tt disks.py}: \\ \\
{\tt import yield\_settings} \\ \\ {\tt VICE} can then be run as normal.  For instructions on how to implement the Gaussian migration scheme, see L. Dubay et al. (submitted).  
Note also that the calculations in this paper tie the yield metallicity-dependence to the scaled oxygen abundance (Equation~\ref{eqn:zoscale}), which requires modifying the VICE source code.  However, we expect only small differences from using the standard VICE procedure of calculating $Z$ from all tracked elements rather than oxygen alone.

Below, we also provide the yield table from \citet{Chieffi2018} for clarity on our calculations regarding the differences between net and gross helium yields.
\setcounter{table}{0}
\renewcommand{\thetable}{A\arabic{table}}

\begin{table*}{}
    \centering
    \caption{Table from \citet{Chieffi2018} reporting the metallicity, birth helium abundance, initial mass, remnant mass, gross explosive yields, gross wind yields, total gross yields, and the calculated total net yields.}
    
    \label{tab:lc18}
    \begin{tabularx}{0.725\textwidth}{r r r r r r r r} \\ \toprule
    \multicolumn{1}{r}{[M/H]} & \multicolumn{1}{r}{$Y$} & \multicolumn{1}{r}{$m$ [\M]} & \multicolumn{1}{r}{$m_{\mathrm{rem}}$ [\M]} & \multicolumn{1}{r}{$m\mathrm{_{He, exp}^{gross}}$ [\M]} & \multicolumn{1}{r}{$m\mathrm{_{He, wind}^{gross}}$ [\M]} & \multicolumn{1}{r}{$m\mathrm{_{He}^{gross}}$ [\M]} & \multicolumn{1}{r}{$m\mathrm{_{He}^{net}}$ [\M]} \\ \midrule
    0 & 0.265 & 13.0 & 1.52 & 4.3131 & 0.3196 & 4.6327 & 1.5905 \\
    0 & 0.265 & 15.0 & 1.63 & 4.6231 & 0.4995 & 5.1226 & 1.5796 \\
    0 & 0.265 & 20.0 & 1.62 & 3.3744 & 3.7908 & 7.1652 & 2.2945 \\
    0 & 0.265 & 25.0 & 1.87 & 2.2278 & 6.1010 & 8.3288 & 2.1994 \\
    0 & 0.265 & 30.0 & 10.8 & 0.0 & 7.3349 & 7.3349 & 2.2469 \\
    0 & 0.265 & 40.0 & 14.1 & 0.0 & 10.9740 & 10.9740 & 4.1105 \\
    0 & 0.265 & 60.0 & 17.0 & 0.0 & 21.6100 & 21.6100 & 10.2150 \\
    0 & 0.265 & 80.0 & 22.7 & 0.0 & 30.0980 & 30.0980 & 14.9135 \\
    0 & 0.265 & 120.0 & 27.9 & 0.0 & 51.9770 & 51.9770 & 27.5705 \\
    -1 & 0.250 & 13.0 & 1.53 & 4.4366 & 0.1280 & 4.5646 & 1.6971 \\
    -1 & 0.250 & 15.0 & 1.69 & 4.5367 & 0.2047 & 4.7414 & 1.4139 \\
    -1 & 0.250 & 20.0 & 1.74 & 6.4833 & 0.4172 & 6.9005 & 2.3355 \\
    -1 & 0.250 & 25.0 & 2.48 & 6.5957 & 1.1494 & 7.7451 & 2.1151 \\
    -1 & 0.250 & 30.0 & 28.3 & 0.0 & 0.4309 & 0.4309 & 0.0059 \\
    -1 & 0.250 & 40.0 & 28.7 & 0.0 & 3.2393 & 3.2393 & 0.4143 \\
    -1 & 0.250 & 60.0 & 42.0 & 0.0 & 6.0919 & 6.0919 & 1.5919 \\
    -1 & 0.250 & 80.0 & 39.9 & 0.0 & 16.9170 & 16.9170 & 6.8920 \\
    -1 & 0.250 & 120.0 & 50.7 & 0.0 & 37.1280 & 37.1280 & 19.8030 \\
    -2 & 0.240 & 13.0 & 1.61 & 4.4375 & 0.0080 & 4.4455 & 1.7119 \\
    -2 & 0.240 & 15.0 & 1.65 & 5.0859 & 0.0502 & 5.1361 & 1.9321 \\
    -2 & 0.240 & 20.0 & 1.76 & 6.6975 & 0.0656 & 6.7631 & 2.3855 \\
    -2 & 0.240 & 25.0 & 2.01 & 8.2445 & 0.0818 & 8.3263 & 2.8087 \\
    -2 & 0.240 & 30.0 & 29.9 & 0.0 & 0.0331 & 0.0331 & 0.0091 \\
    -2 & 0.240 & 40.0 & 39.7 & 0.0 & 0.0606 & 0.0606 & -0.0114 \\
    -2 & 0.240 & 60.0 & 59.6 & 0.0 & 0.0918 & 0.0918 & -0.0042 \\
    -2 & 0.240 & 80.0 & 78.6 & 0.0 & 0.3365 & 0.3365 & 0.0005 \\
    -2 & 0.240 & 120.0 & 103.0 & 0.0 & 4.5422 & 4.5422 & 0.4622 \\
    -3 & 0.240 & 13.0 & 1.61 & 4.4429 & 0.0051 & 4.4480 & 1.7144 \\
    -3 & 0.240 & 15.0 & 1.8 & 4.6174 & 0.0113 & 4.6287 & 1.4607 \\
    -3 & 0.240 & 20.0 & 1.8 & 6.6192 & 0.0494 & 6.6686 & 2.3006 \\
    -3 & 0.240 & 25.0 & 2.15 & 7.9727 & 0.0871 & 8.0598 & 2.5758 \\
    -3 & 0.240 & 30.0 & 30.0 & 0.0 & 0.0048 & 0.00481 & 0.00481 \\
    -3 & 0.240 & 40.0 & 40.0 & 0.0 & 0.0077 & 0.0077 & 0.0077 \\
    -3 & 0.240 & 60.0 & 59.9 & 0.0 & 0.0137 & 0.0137 & -0.0103 \\
    -3 & 0.240 & 80.0 & 79.9 & 0.0 & 0.0237 & 0.0237 & -0.0003 \\
    -3 & 0.240 & 120.0 & 120.0 & 0.0 & 0.0819 & 0.0819 & 0.0819 \\ \bottomrule
    
\end{tabularx}
\end{table*}